\documentclass{article}

\usepackage{PRIMEarxiv}

\usepackage[utf8]{inputenc} 
\usepackage[T1]{fontenc}    
\usepackage{hyperref}       
\usepackage{url}            
\usepackage{amsmath,xparse}

\usepackage{booktabs}       
\usepackage{amsfonts}       
\usepackage{nicefrac}       
\usepackage{microtype}      
\usepackage{lipsum}
\usepackage{fancyhdr}       
\usepackage{graphicx,caption}
\usepackage{subfig}
\usepackage{tabularx}   
\graphicspath{{media/}}     
\usepackage{url}
\makeatletter
\g@addto@macro{\UrlBreaks}{\UrlOrds}
\makeatother

\title{Analyzing and visualizing polarization and balance with signed networks: the U.S. Congress case study}

\author{
  Arthur Capozzi\\
  Department of Computer Science\\
  Università degli Studi di Torino \\
  Via Pessinetto, 12, 10149, Torino, Italy\\
  \texttt{Corresponding author: arthurtomasedward.capozzi@unito.it} \\
  \And
  Alfonso Semeraro\\
  Department of Computer Science\\
  Università degli Studi di Torino \\
  Via Pessinetto, 12, 10149, Torino, Italy\\
   \And
  Giancarlo Ruffo\\
  Department of Science and Technological Innovation (DISIT)\\
  Università del Piemonte Orientale ``A. Avogadro''\\
  Viale Teresa Michel, 11, 15121 Alessandria, Italy\\
}

\begin{document}
\maketitle

\begin{abstract}
Signed networks and balance theory provide a natural setting for real-world scenarios that show polarization dynamics, positive/negative relationships, and political partisanship~\cite{plosone}. For example, they have been proven effective in studying the increasing polarization of the votes in the two chambers of the U.S. Congress from World War II on~\cite{plosone, aref2020}.

To provide further insights into this particular case study, we propose the application of a pipeline to analyze and visualize a signed graph’s configuration based on the exploitation of the corresponding Laplacian matrix’ spectral properties~\cite{Galimberti}.
The overall methodology is comparable with others based on the frustration index~\cite{Aref2019}, but it has at least two main advantages: first, it requires a much lower computational cost; second, it allows for a quantitative and visual assessment of how arbitrarily small subgraphs (even single nodes) contribute to the overall balance (or unbalance) of the network.

The proposed pipeline allows the exploration of polarization dynamics shown by the U.S. Congress from 1945 to 2020 at different resolution scales. In fact, we are able to spot and point out the influence of some (groups of) congressmen in the overall balance, as well as to observe and explore polarization’s evolution of both chambers across the years.
\end{abstract}


\section{Introduction}
Studying opinion and political polarization dynamics in social networks is considered crucial to understand how individuals that belong to different communities can be exposed to a plurality of information and beliefs~\cite{Conover_Ratkiewicz_Francisco_Goncalves_Menczer_Flammini_2021}. The consequences of strong polarization, such as the emergence of the so-called ``echo-chambers''~\cite{sunstein2001} and increased partisanship in politics~\cite{plosone}, can represent a risk for modern democracies. They can lead to ``filter bubbles'' undermining social media recommendation algorithms~\cite{pariser2011}, accelerate the spreading of misinformation~\cite{lazer2018}, among other things. There is a general agreement that polarization is a negative consequence of network's homophily, and explicit relationships between individuals have been intensively used to study related phenomena in complex social systems. Friendship is usually represented by means of positive relationships, even if in many different domains it is important to consider also negative ties to better understand polarization dynamics~\cite{Keuchenius2021}. 

Signed networks can effectively represent patterns of collaboration, friendship, or membership in a group, by means of positive edges, as well as explicit disagreements, different opinions, antagonisms, by means of negative edges.

In this type of networks, the coexistence of two forces inevitably leads to the formation of cohesive groups of individuals, also called clusters. A \textit{balanced network} represents a perfectly polarized system, where there are only positive relationships between nodes of the same group and negative edges between nodes belonging to different clusters. Perfectly balanced systems rarely manifest themselves in practice; however, we can have several shades of unbalance. 

In many analyses, we may want to check both the presence of homophily/heterophily due to the node attribute and how far a network is from being balanced: these measures could help to understand the polarization dynamics of the system from different points of view. Political systems, as roll call votes in the U.S. Congress, are paradigmatic: we can label nodes as belonging to a party (e.g., Democrats or Republicans), and we expect that positive links are more frequent than negative ones between intra parties representatives, as well as the opposite between inter parties nodes. Observing how balance changes over time can help us understand polarization dynamics; monitoring homophily may support the analyst to reveal signals of partisanship or anti-partisanship behaviors among (group of) nodes in the network.

\subsection{Our contribution}
Although there are several algorithms and measures in the literature that would support signed network analysis, a pipeline that combines these techniques is still missing.

In this paper, we present an analytical pipeline that leverages techniques and metrics from the literature to efficiently analyze and visualize signed networks in terms of both homophily and balance. We present the potential of this pipeline to dissect parts of the network and assess their contribution to the overall structural balance, allowing for a fine-grained analysis of the actors (and the parties they belong to) that make the network less balanced and unstable in its configuration.

We showcase this analytical pipeline by studying the polarization of the votes in the two chambers of the U.S. Congress from 1945 to 2020, where nodes have explicit attributes denoting their membership to a specific party. The results are consistent with others in the literature, and a fine-grained analysis of Senator Manchin's political transformation demonstrates the new potential of this pipeline.

The measures adopted to support the analysis and visualization of the signed graph are based on the spectral analysis of the Laplacian matrix.
The computation of the smallest eigenvalue $\lambda_n$ can be used to understand whether the network is balanced and, if not, how far it is from being balanced. We compare the results and the computational cost to another metric based on the frustration index~\cite{Aref2019}. Furthermore, the eigenvector $v_n$ corresponding to the smallest eigenvalue $\lambda_n$, as suggested by Kunegis et al.~\cite{Kunegis2} and Galimberti et. al.~\cite{Galimberti}, can be used to i) study the influence of arbitrarily large subgroups of nodes on the overall balance, and ii) display nodes in a Cartesian plane depending on their individual contribution to the network's balance.

Given the great opportunities of applying signed networks in real-world scenarios, a Python library containing the analytics and visualizations underlying the pipeline presented in this study is being built and released. In this library, called Sygno\footnote{\url{https://github.com/alfonsosemeraro/draw_signed_networkx}}, we intend to implement further functions to support the study of signed networks and balance theory.

Related work is reviewed in Section~\ref{sec:related}, while in Section~\ref{sec:methods}: i) we summarize some basic definitions of signed networks, spectral analysis, and structural balance; ii) we give a mathematical definition of the metrics used in this paper to compute the structural balance. In Section~\ref{sec:netcreation}, we describe the construction of signed networks starting from the voting data of the U.S. Congress.

In Section \ref{sec:results}, we present a comparative analysis of two structural balance computation metrics; we show how the spectral analysis of the Laplacian matrix and the data visualization can support the study of the influence of individual nodes on the network balance. We conclude by presenting a computational cost comparison of structural balancing computation metrics.

\section{Related works}
\label{sec:related}

\subsection{Signed networks}
\label{sec:signednets}

Network science has many applications in social media and social networks analysis~\cite{Knoke2008,Aggarwal}. Networks that only represent relationships of friendships or following in social media can be seen as signed networks with only positive edges. Many scholars have studied how to deal with the increased complexity of the models that also represent the relations of antagonism through negative edges, Tang et al.~\cite{Tang2017} describe the process of mining social media data to build signed networks based models. The increase in complexity makes modeling the network more difficult, but it also provides many opportunities for further analyses. 

Balance theory is central to many studies involving signed networks. The theory's fundamental observation is that three nodes linked to each other can show both stable and unstable configurations. For instance, the enmity between two nodes in a friendly relationship with a third node is a source of contrast, and the triangle is defined unbalanced. Thus, the \emph{structural balance} of a signed network is the overall property of the network to be built upon stable triangles. In social networks, the concept of balance was first formulated by Heider in 1946~\cite{Heider}. He theorized that the relationships that form the structure of social networks may be characterized by the dualism of friendship and hostility. Heider's studies influenced the social psychology of the years to come, but it was only in 1956 that the balance was modeled in terms of signed graphs~\cite{Cart}; in fact, Cartwright and Harary's intent was to mathematically formalize balance theory and to extend it to other domains, such as communication networks, power systems, sociometric structures, systems of orientations, or neural networks.

Nowadays many applications of this theory to signed networks can be found in the literature; e.g., it has been used to analyze the interactions and polarization of users in social media~\cite{Leskovec,Kunegis}, or to study international relations between countries~\cite{Doreian}, or biotic interactions in real ecosystems~\cite{Saiz}.

In particular, Leskovec et al.~\cite{Leskovec2010a} tried to evaluate social-psychological theories on the balancing of social systems by modeling large-scale datasets with signed networks. They also proved that positive link prediction can be improved by using both positive and negative edges~\cite{Leskovec2010b}. Other studies have focused on the application of signed networks with the aim of improving the performance of recommendation systems~\cite{Victor2011,Ma2009}.

Some studies have modeled recommendation systems~\cite{Chen2020} or social networks~\cite{Hua_Liu2020} with time-varying signed networks. In fact, real-world signed networks can change over time in many aspects, such as structure, node clustering, or structural balance.

Another relevant field of application of signed networks concerns the study of the voting behavior of politicians and voters. Porco et al.~\cite{PorcoA2015} tried to predict polarity and voting behavior by using a recommendation system-based approach. Intal and Yasseri~\cite{Intal2021} analyzed the individual interactions of House of Commons Members of Parliament (MPs) in the voting process. The purpose of their study was to predict how MPs would have voted in a forthcoming Brexit deal. They computed pairwise similarity scores and calculated ``rebellion'' metrics based on eigenvector centrality obtaining an accuracy of 90$\%$ in the prediction task. Eigenvector centrality has been shown to be an effective method for quantifying the power and overall strength of a node in a signed network~\cite{Bonacich2007}.

Several aspects of American politics and the U.S. Congress have been the subject of network science studies and analyses. In 2015, Andris et al.~\cite{plosone} analyzed the evolution of partisanship in the United States Congress from 1949 to 2012. From roll-call voting they have created an undirected network of over 5 million pairs of representatives, and showed that partisanship has increased exponentially over the past 60 years.
In 2020, Neal~\cite{neal2020} analyzed the polarization in the U.S. Congress by applying the Stochastic Degree Sequence Model and creating signed networks of political relationships among legislators. An increase in polarization in the U.S. Congress has been shown by using the triangle index, a metric for computing partial structural balance of signed networks.

In 2021 Aref and Neal~\cite{aref2020} studied polarization on political networks proposing new models for optimal clustering of signed networks. They have confirmed the increase in polarization in the US Congress since 1979. They also found that partisanship may positively influence the effectiveness of passing bills. 
In their study, they used two metrics for measuring partial balance: triangle index, i.e., the fraction of positive cycles of length 3, and the frustration index, based on the minimum number of edges that should be removed to get a balanced network.

Similar analyzes regarding political votes and polarization, involve among others the study of the European Parliament~\cite{Arinik2017,Mendonca2015}, the U.S. Congress~\cite{Waugh2009, Dana2013} and the Brazilian House of Representatives~\cite{Ferreira2018}.

\subsection{Compute structural balance}
\label{sec:computebalance}
The first formal definition of structural balance in terms of signed graphs is due to Cartwright and Harary~\cite{Cart}: a signed network $\Gamma$ is balanced if and only if its nodes can be partitioned into two disjoint subsets $V_{1}$, $V_{2}$ in such a way that each positive edge of $\Gamma$ joins two nodes of the same subset and each negative edge joins two nodes of different subsets.

It is possible to determine whether a signed network is balanced or not in polynomial time~\cite{Facchetti}. When a signed network is not balanced, it is possible to compute the partial structural balance, which is the distance of the network from being balanced. Calculating how far a network is from being balanced is much more complex. Approaches based on the computation of the least number of edges that must be dropped to obtain a balanced network, as the frustration index, are nondeterministic polynomial-time hard problems~\cite{Facchetti}. Other approaches are based on the spectral analysis of the Laplacian matrix of the signed graph, and have a complexity of $O(n^2)$, like the algebraic conflict.

Some of the most common approaches are walk-based~\cite{Estrada,Singh_2017,Kirkley}. A walk of length $k$ in $\Gamma$ is a sequence of connected nodes $v_0, \ldots, v_k$. The sign of the walk is the product of the signs of the edges that connect the nodes $v_0, \ldots, v_k$~\cite{Zaslavsky2013MatricesIT}. The walk is defined as closed (or as a cycle) if $v_0 = v_k$ and if its length is equal to $k$. A signed network is walk-balanced if every closed walk is positive~\cite{Estrada}. Hence, a network is balanced if every closed walk has an even number of negative edges~\cite{Facchetti}.

\subsection{Visualize structural balance}
\label{sec:introvizstructbalance}
Many network visualization techniques are proposed in the literature, but very few approaches are designed to visualize specifically signed networks. Kunegis et al.~\cite{Kunegis2} showed how the properties of spectral analysis can be used to draw signed graphs. In order to obtain an embedding of the nodes of a signed graph into a plane, they used the eigenvectors of the two smallest eigenvalues as coordinates for drawing the nodes. Using this method, when the graph is balanced ($\lambda_{|V|} = 0$), all nodes are placed on two parallel lines, reflecting the perfect bi-partitioning present in the graph.

In our study we use another spectral analysis based algorithm called \textit{Structural-balance-viz}~\cite{Galimberti}. If compared with a generic force-based graph drawing layout where edges can be shown as positive/negative (for example as solid/dashed lines), this visualization scheme  has multiple advantages: i) it shows clearly whether the network is balanced or not; ii) it allows to visually represent the partial balance; iii) it is capable of providing information on the contribution of individual nodes to the balance of the entire network; iv) even a slight change to the graph would correspond to a visual modification, allowing to spot and compare differences between different graphs. In Section~\ref{ssec:vizcongress} we provide an accurate description of this method and we apply it to the U.S. Congress roll call votes dataset.

\section{Methods}
\label{sec:methods}

In this section we define the partial balance indices that we compute and compare in section~\ref{sec:results} to analyze the temporal dynamics of political polarization from 1941 to 2020 in the U.S. Congress. 

\subsection{Notations and metrics}

Let $\Gamma = (G, \sigma)$ be an undirected signed graph, where $G = (V(G), E(G))$ is a graph, and $\sigma$ is the sign function $\sigma: E(G) \rightarrow \{{-1}, {+1}\}$. We define $E_{+}$ as the set of positive edges, and $E_{-}$ as the set of negative edges. $A$ is the signed adjacency matrix of $\Gamma$, i.e., for each pair of nodes $u, v \in V$, $A[u, v] = 1 $ if $ (u, v) \in E_{+}, A[u, v] = -1 $ if $ (u, v) \in E_{-}$. Let $\bar{D} = diag(\bar{d}_{u1},\dots,\bar{d}_{u|V|})$ be the signed degree matrix of $\Gamma$, where $\bar{d}_u = \sum_{v\in V}|A[u,v]|$.
Finally, we define the signed Laplacian matrix $\bar{L} \in \mathbb{R}^{V \times V}$ of $\Gamma$ as:
\begin{equation}
        \bar{L}(\Gamma) = \bar{D} - A
\end{equation}
The Laplacian matrix spectrum captures some characteristics of the relationship between the nodes. It is composed by the $|V|$ Laplacian eigenvalues $\lambda_1 \geq \lambda_2 \geq \ldots \geq \lambda_{|V|}$. Laplacian eigenvalues are all real and non-negative, and contained in the interval $[0, min\{|V|, 2d_{max}\}]$, where $d_{max}$ is the maximum degree of $\Gamma$~\cite{DAS2004715}.
The second smallest eigenvalue $\lambda_{|V|-1}$ is equal to zero only if the graph is disconnected. If $\lambda_{|V|-1} > 0$, then its multiplicity indicates how many connected components exist in $\Gamma$~\cite{Marsden2013EIGENVALUESOT}. Furthermore, as we report below, the smallest eigenvalue captures some structural balance properties of signed graphs.

In this study we apply and compare the following two approaches for computing partial structural balance:

\begin{itemize}
	\item Frustration index ($\epsilon(\Gamma)$): it is equal to the absolute value of the \textit{minimum detection set} ($|E^*|$), that is the lowest number of edges to be deleted from a signed network to obtain a balanced network. Computing this metric is very expensive from a computational point of view: in the worst case, when all edges are negative, the frustration index is NP-hard (it is equivalent to the MAX-CUT problem~\cite{barahona1983}). In this study we use the \textit{normalized frustration index} suggested in Aref~\cite{Aref2019}:
	\begin{equation}
	    \label{eq:frustration}
        F(\Gamma) = 1 - \frac{\epsilon(\Gamma)}{m/2}
    \end{equation}
    Where $m$ is the number of edges of $\Gamma$.
	\item Algebraic conflict ($\lambda_{|V|}(\Gamma)$): it is the smallest eigenvalue of the signed Laplacian matrix. In this study we use the \textit{normalized algebraic conflict} suggested by Aref~\cite{Aref2019}:
	\begin{equation}
	    \label{eq:algebconf}
        A(\Gamma) = 1 - \frac{\lambda_{|V|}(\Gamma)}{\bar{d}_{max} - 1},\;\;\;\;\;\;\;\;\; \bar{d}_{max} = \underset{(u, v)\in E}{max} (d_u + d_v)/2
    \end{equation}
\end{itemize}

In positively weighted graphs, when the smallest eigenvalue $\lambda_{|V|}(\Gamma)$ is nonzero, the corresponding eigenvector $u_{|V|}$ provides the embedding coordinates of each vertex~\cite{Belkin}. In signed networks, the sign of the elements in $u_{|V|}$ can be used for partitioning the nodes into two clusters~\cite{ColemanClustering, Galimberti}, and the value can be used as coordinates~\cite{Kunegis}.

The normalized frustration index equals to one if the signed network is perfectly balanced; otherwise, a value less than one measures the partial structural balance.

If a signed graph is balanced, the corresponding Laplacian matrix is proved to be positive semidefinite (all its eigenvalues are non-negative), and $\lambda_{|V|}(\Gamma) = \epsilon(\Gamma) = 0$~\cite{Kunegis, BELARDO2014133}. If a signed graph is unbalanced, the corresponding Laplacian matrix is proved to be positive-definite, and an algebraic conflict greater than zero estimates how much the signed network is far from being balanced~\cite{Hou2005BoundsFT}. Furthermore, it has been proved that $\lambda_n(\Gamma) \leq \epsilon(\Gamma)$~\cite{BELARDO2014133}.

\subsection{Spectral analysis and structural balance}

\begin{figure}[!tbp]
  \centering
  \captionsetup{width=.92\linewidth}
  \subfloat{\includegraphics[width=0.3\textwidth]{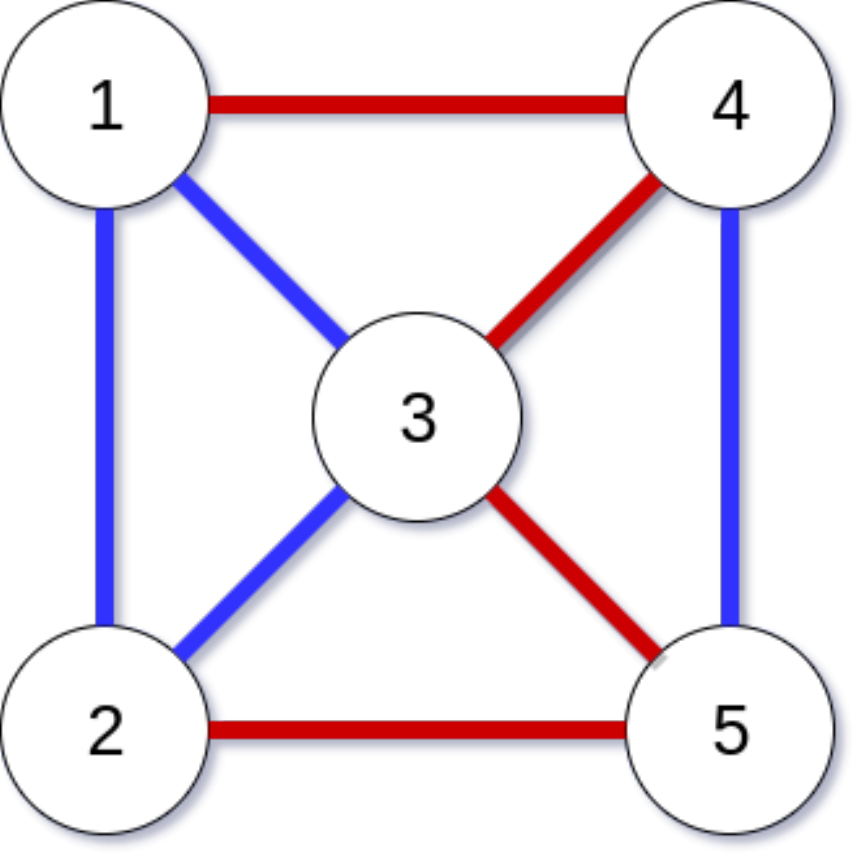}}
  \hfill
  \subfloat{\includegraphics[width=0.5\textwidth]{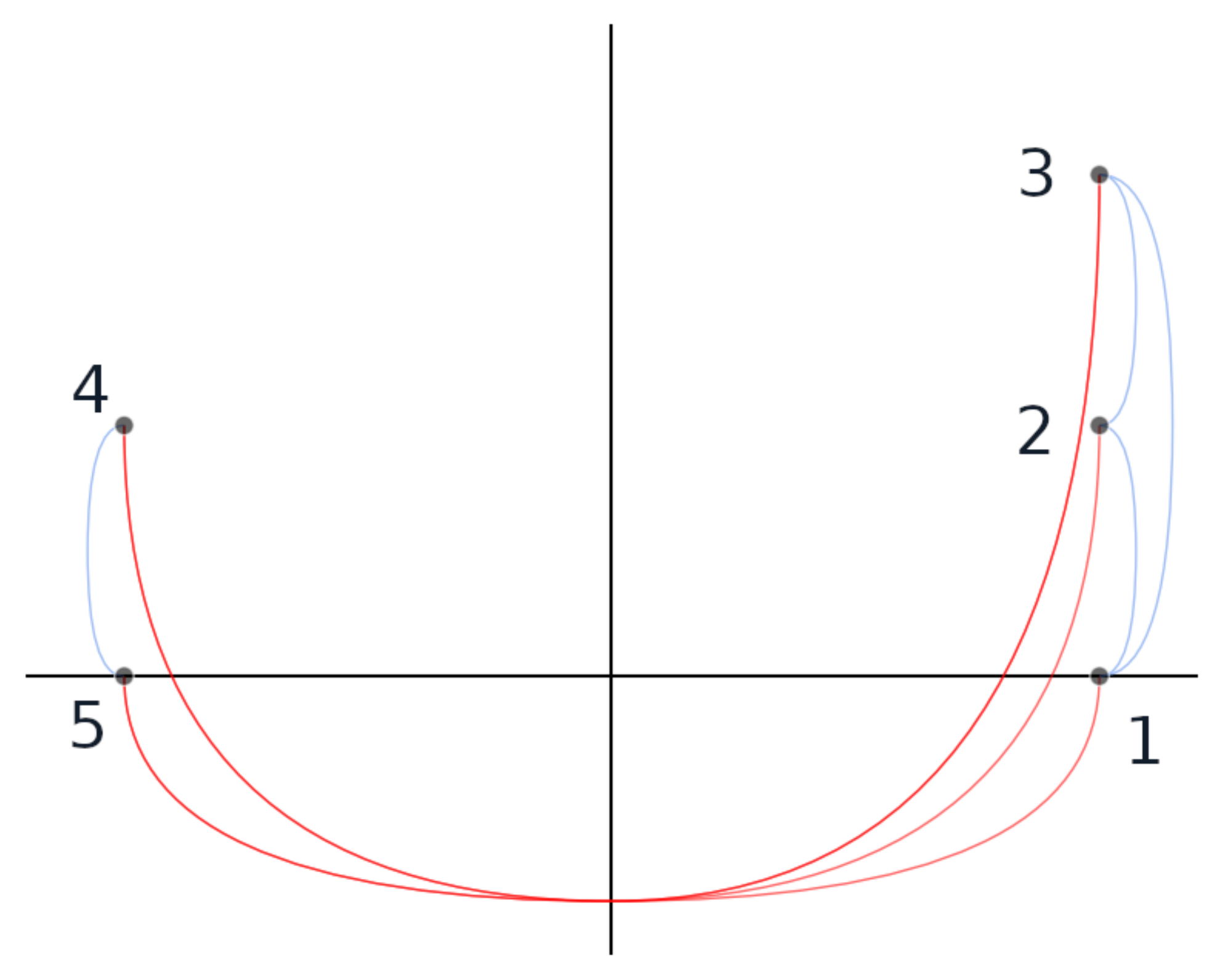}}
  \caption{A balanced signed network is represented on the left. The first cluster is composed by nodes 1, 2, and 3. The second cluster consist of nodes 4 and 5. On the right there is the visualization of the same network created with the \textit{Structural-balance-viz} algorithm. In both representations, positive edges are blue, and negative edges are red.}
  \label{fig:5_nodes_balanced}
\end{figure}

A balanced signed network is shown in Figure~\ref{fig:5_nodes_balanced}. The vertices can be divided into two subsets, with only negative edges between them and only positive edges within them. The first subset consists of vertices 1, 2 and 3, the second of vertices 4 and 5. The equation in~\ref{eq:network_d_a_balanced} shows the computation of $L$ as $D - A$. The vector of the eigenvalues of $L$, and the eigenvector $u_{|V|}$ corresponding to the smallest eigenvalue are shown in equation~\ref{eq:L_balanced}. The smallest eigenvalue ($\lambda_{|V|}(\Gamma)$) of the signed Laplacian matrix $L$ is equal to zero and this confirms that the network is perfectly balanced. The equation shows also the computation of the partial structural balance using the normalized algebraic conflict (Eq.~\ref{eq:algebconf}). The network is balanced, so $A(\Gamma)$ is equal to 1.

\begin{equation}
\begin{bmatrix}
  3. & 0. & 0. & 0. & 0.\\
  0. & 3. & 0. & 0. & 0.\\
  0. & 0. & 4. & 0. & 0.\\
  0. & 0. & 0. & 3. & 0.\\
  0. & 0. & 0. & 0. & 3.\\
\end{bmatrix}
-
\begin{bmatrix}
  0 & 1 & 1 & -1 & 0\\
  1 & 0 & 1 & 0 & -1\\
  1 & 1 & 0 & -1 & -1\\
  -1 & 0 & -1 & 0 & 1\\
  0 & -1 & -1 & 1 & 0\\
\end{bmatrix}
=
\begin{bmatrix}
  3. & -1. & -1. & 1. & 0.\\
  -1. & 3. & -1. & 0. & 1.\\
  -1. & -1. & 4. & 1. & 1.\\
  1. & 0. & 1. & 3. & -1.\\
  0. & 1. & 1. & -1. & 3.\\
\end{bmatrix}
\label{eq:network_d_a_balanced}
\end{equation}

\begin{equation}
\lambda(\Gamma)=
\begin{bmatrix}
  0. & 3. & 5. & 3. & 5.\\
\end{bmatrix}
, u_{|V|}=
\begin{bmatrix}
  0.4472\\
  0.4472\\
  0.4472\\
  -0.4472\\
  -0.4472\\
\end{bmatrix}
\label{eq:L_balanced}
\end{equation}

\begin{equation}
    A(\Gamma) = 1 - \frac{0}{\bar{d}_{max} - 1} = 1,\;\;\;\;\;\;\;\;\; \bar{d}_{max} = 3.5
\label{eq:plot_edo_balanced}
\end{equation}

The visualization on the right of Figure~\ref{fig:5_nodes_balanced} is created with the \textit{Structural-balance-viz} (SBV) algorithm~\cite{Galimberti}. SBV creates reproducible visualizations that show the degree of partial network balancing and clearly identify clusters of nodes. The eigenvector $u_{|V|}$ is used to show the contribution of each single node to the structural balance of the network.
In \textit{Structural-balance-viz} the nodes of the network are drawn on a Cartesian plane and the coordinates are calculated as it follows:
\begin{itemize}
    \item x of node u is equivalent to the element of the eigenvector corresponding to u.
    \item y is calculated in such a way as not to overlap the nodes with the same x coordinate.
\end{itemize}

\textit{Structural-balance-viz} shows two types of edges:
\begin{itemize}
    \item \textit{Frustrated edges} decrease the structural balance of the network. These edges can be negative (in red) if they connect nodes in the same cluster (i.e., in the same quadrant of the Cartesian plane); or positive (in blue) if they connect two nodes of different clusters (i.e., in different quadrants of the Cartesian plane).
    \item \textit{Non-frustrated edges} represent an agreement between two nodes of the same cluster (positive edge), or a disagreement between two nodes of different clusters (negative edge). Positive and negative non-frustrated edges increase the structural balancing of the network.
\end{itemize}

The computational time required to execute the SBV algorithm is the same as that required to compute the algebraic conflict.

Since the network in Figure~\ref{fig:5_nodes_balanced} is perfectly balanced, the nodes are represented in two distinct clusters and there are no positive edges between the two clusters nor negative edges within the clusters, so there are no frustrated edges.


\begin{figure}[!tbh]
  \centering
  \captionsetup{width=.92\linewidth}
  \subfloat{\includegraphics[width=0.3\textwidth]{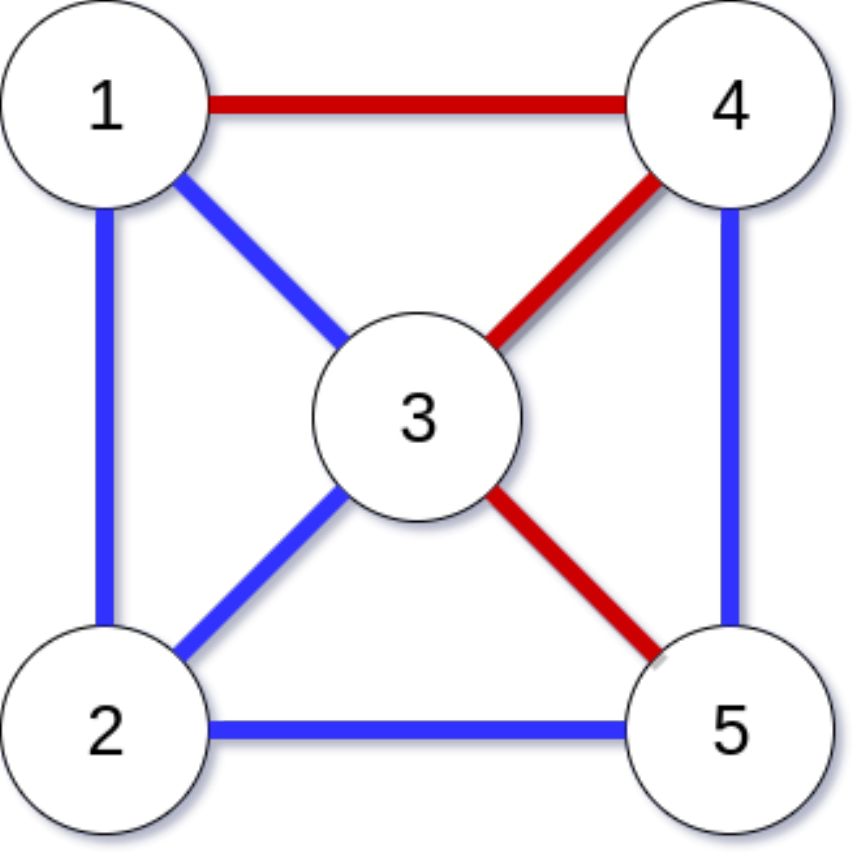}}
  \hfill
  \subfloat{\includegraphics[width=0.5\textwidth]{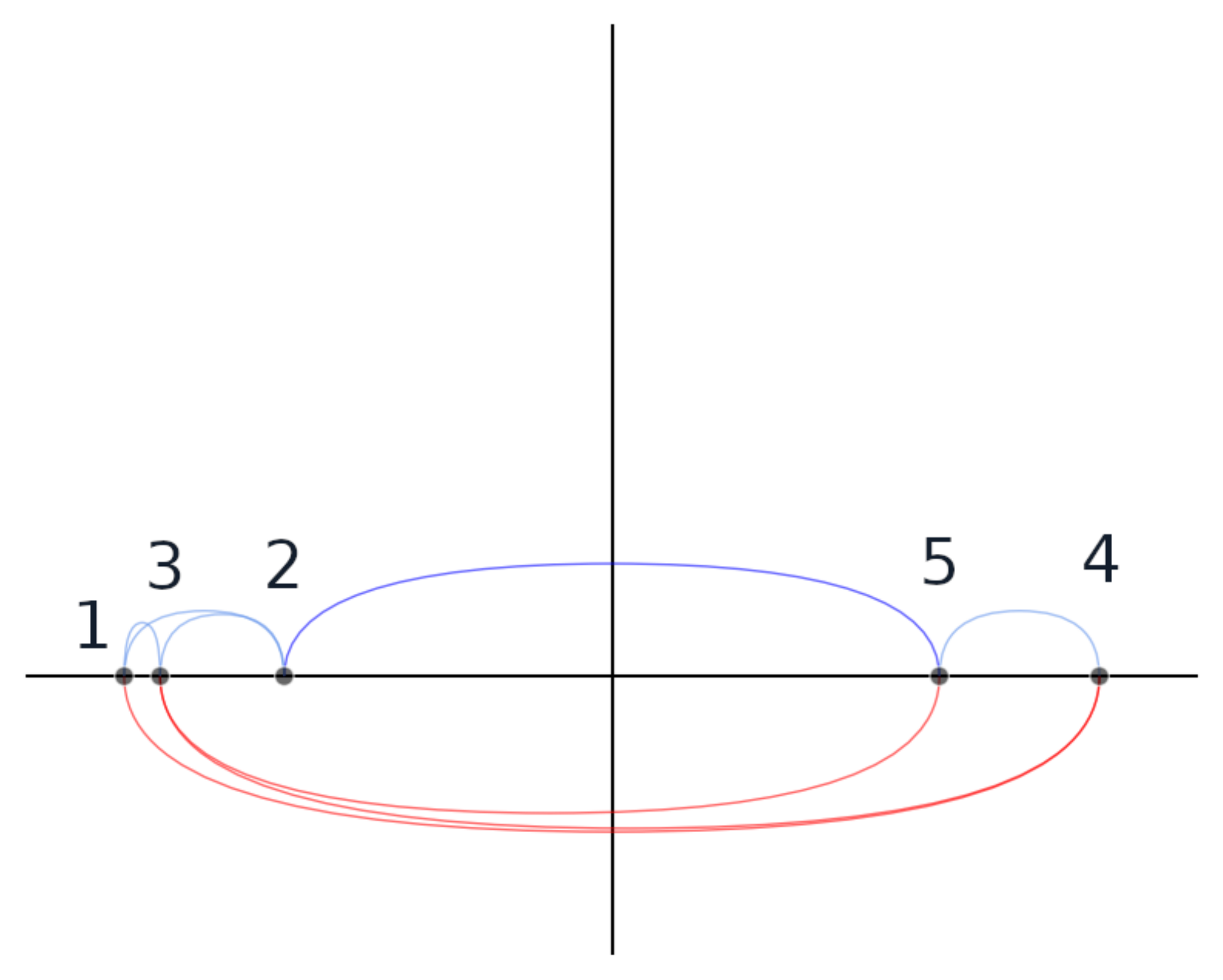}}
  \caption{An unbalanced signed network is represented on the left. This network is obtained by adding to the network in Fgure~\ref{fig:5_nodes_balanced} a positive edge between nodes 2 and 5. On the right there is the visualization of the same network created with the \textit{Structural-balance-viz} algorithm.}
  \label{fig:5_nodes_unbalanced}
  
\end{figure}

In Figure~\ref{fig:5_nodes_unbalanced} we show a variation of the same network introduced above, but with a negative edge between vertices 2 and 5. In this case, the network is no longer balanced as shown in Eq.~\ref{eq:edo_plot_unbalanced} by a value of $A(\Gamma)$ less than 1. In Figure~\ref{fig:5_nodes_unbalanced} it is shown the visualization created with the \textit{Structural-balance-viz} algorithm. The positive frustrated edge between 2 and 5 is colored in blue and connects the two clusters, and it actually crosses the y-axis, instead of increasing inter-group clustering. Even a small variation in the network (and its balance) produces a significant difference in the corresponding visualization, making comparisons self-explanatory.

\begin{equation}
\lambda(\Gamma)=
\begin{bmatrix}
  0.5505 & 1.5857 & 4. & 4.4142 & 5.4494\\
\end{bmatrix}
, u_{|V|}=
\begin{bmatrix}
  -0.5411\\
  -0.2979\\
  -0.4865\\
  0.5411\\
  0.2979\\
\end{bmatrix}
\label{eq:unbalanced_l_u}
\end{equation}

\begin{equation}
    A(\Gamma) = 1 - \frac{0.5505}{\bar{d}_{max} - 1} = 0.7797,\;\;\;\;\;\;\;\;\; \bar{d}_{max} = 3.5
\label{eq:edo_plot_unbalanced}
\end{equation}

This visualization paradigm allows for spotting intra-groups and inter-groups dynamics as well; for example, nodes' attributes could be represented by means of different colors or shapes. Intra-groups and inter-groups edges can be compared to each other to measure homophily. For example, if negative frustrated edges are established between nodes belonging to the same group, or, conversely, if positive frustrated edges connect nodes belonging to different groups, we can detect signals of inverse homophily that contribute to reduce the polarization in the system, making the network less balanced. In the next two sections we show how we can consider inter-groups and intra-groups positive/negative edges to observe both homophily and balance.

\section{Creation of the Congress network}
\label{sec:netcreation}
In this study, we use congressional roll call votes to analyze the dynamics of political polarization in the U.S. Congress from 1945 (77th Congress) to 2020 (117th Congress). The polarization of the U.S. Congress has been extensively studied in the literature, both through the paradigm of unsigned networks~\cite{plosone} and through the paradigm of signed networks~\cite{neal2020, aref2020}. Our paper proposes a quantitative and visual analysis at a new level of detail, namely of individual nodes. Furthermore, using political voting as an emblematic case study, we show how it is possible to analyze polarization in terms of both homophily and structural balance. In this section, we present the vote processing and network construction pipeline for both the House of Representatives and U.S. Senate.

The data are available by the Voteview Project\footnote{\url{https://voteview.com/}}. In a roll call vote, a representative can vote 'yay' or 'nay', and we consider non-attendances as 'nay'. A complete signed network ($G_v$) is created for each vote. A node in the network represents a politician, and there is a positive edge between two nodes if they voted similarly, otherwise there is a negative edge. Trivially, the network representing a single roll call vote is always balanced by construction, since the 'yay' voters are grouped together in one cluster and the 'nay' voters are grouped together in another cluster. Within a cluster, there are only positive edges. Between the clusters there are only negative edges.

For each two-year Congress, we collapse the networks representing each vote into a single weighted network ($G_c$), where the weight of an edge represents the number of collapsed edges. The sum of the weights of the edges connecting two politicians is equal to the number of times they have participated in the same roll call votes. Furthermore, in a similar way to~\cite{plosone}, an edge is labeled as ``inter-party'' if it connects two representatives of two different parties, and as ``intra-party'' if they are from the same party.

$G_c$ is not always a complete network: in fact, if two politicians have never taken part in the same vote, then there is neither a positive nor a negative edge between them. Such cases occur, for example, when a politician joins the House of Representatives or the U.S. Senate at the end of a Congress.

For each pair of nodes of $G_c$ we compute the percentage of votes in agreement ($p_+$) as the weight of the positive edges divided by the sum of the weights of the positive and negative edges (see equations~\ref{eqn:per_agreement}). The same process is used to compute the percentage of disagreements ($p_-$). 
\begin{equation}
\label{eqn:per_agreement}
    \begin{split}
        p_+ = e_+/ (e_+ + e_-)\\
        p_- = e_-/ (e_+ + e_-)
    \end{split}
\end{equation}
Where $e_+$ and $e_-$ are respectively the positive and negative edges between two nodes. For each pair of nodes we keep only the edge with the highest percentage $p_+$ or $p_-$.

\begin{figure}[!tbp]
  \centering
  \captionsetup{width=.92\linewidth}
  \subfloat{\includegraphics[width=0.48\textwidth]{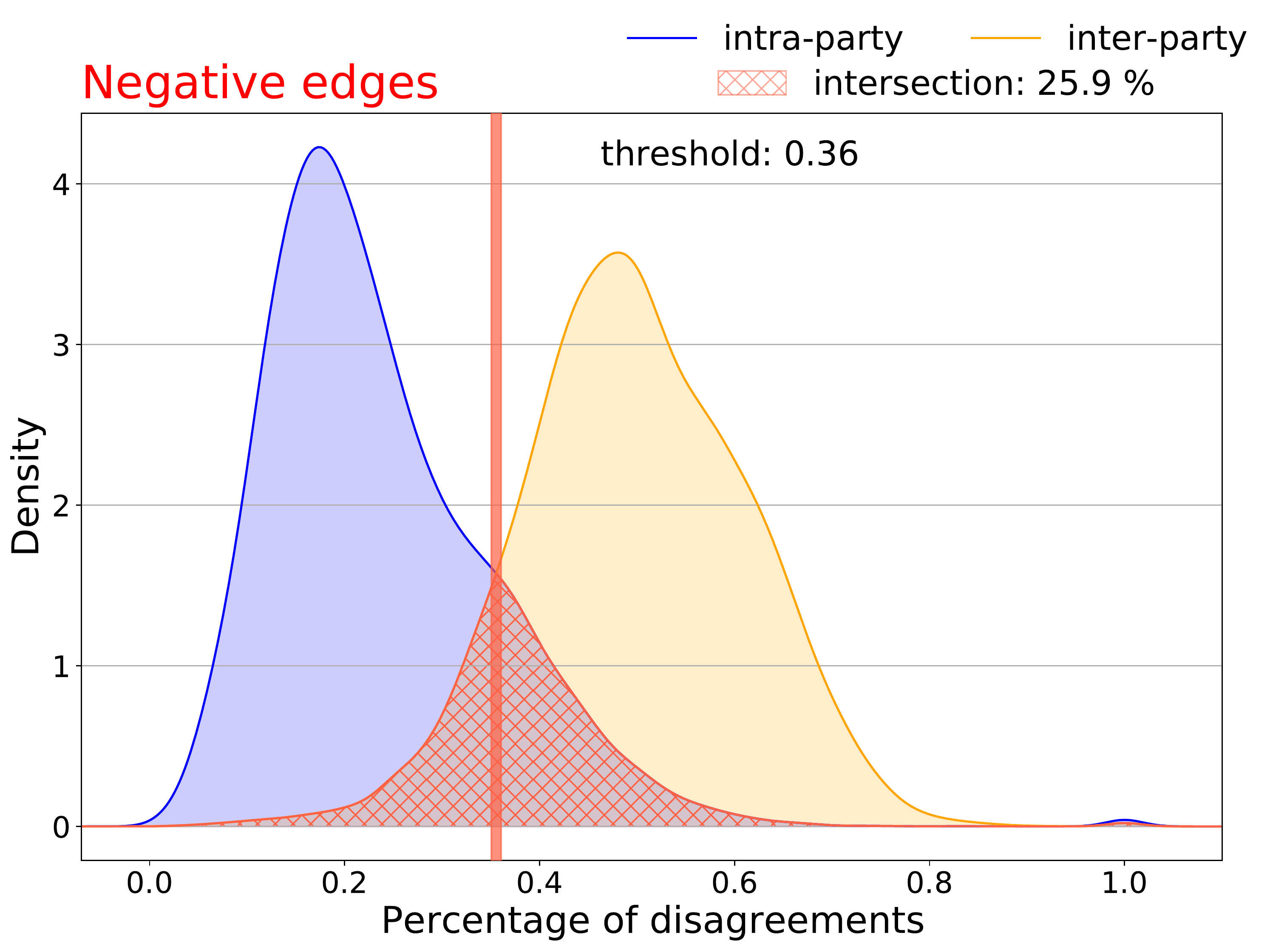}}
  \subfloat{\includegraphics[width=0.48\textwidth]{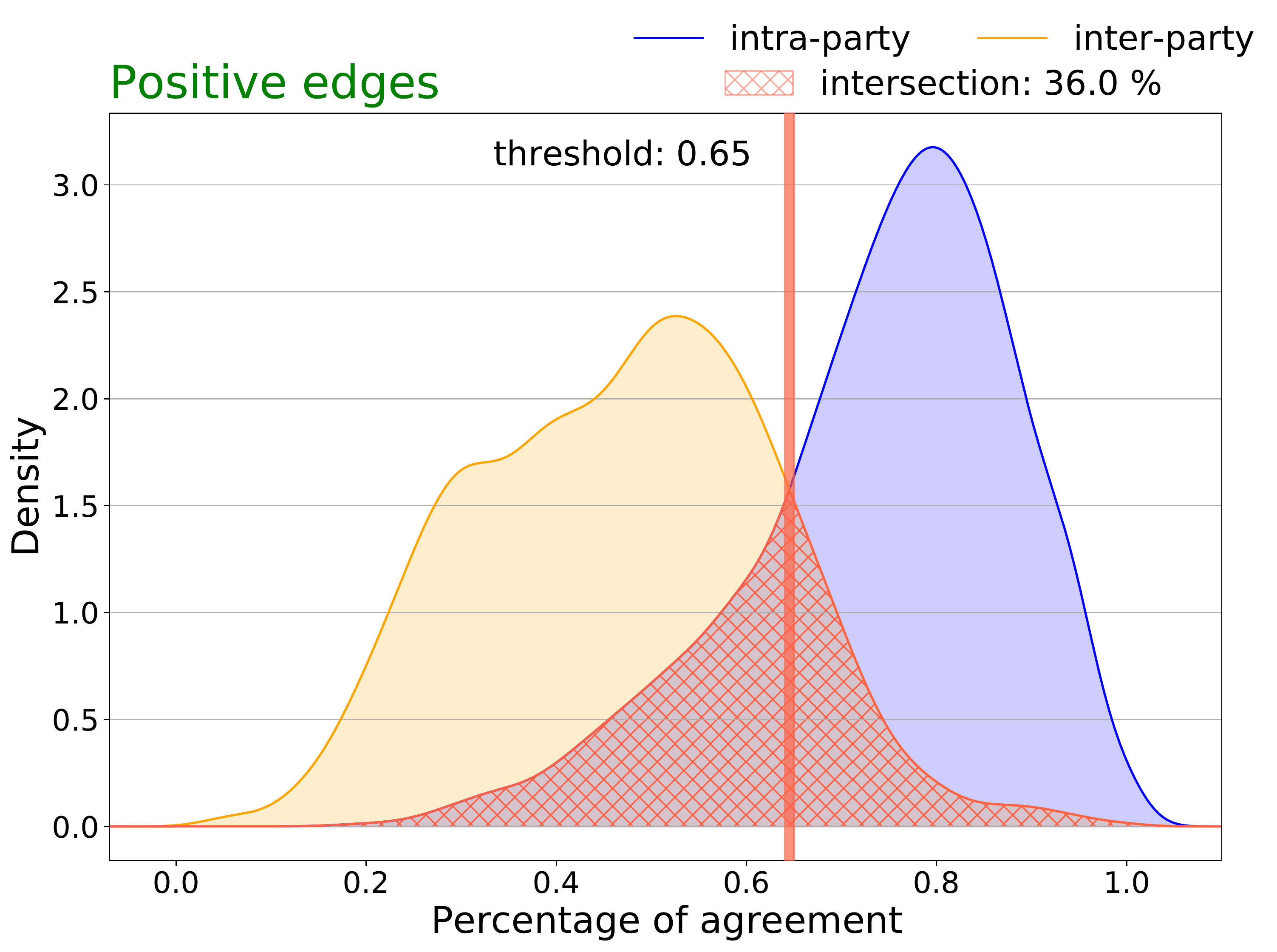}}
  \caption{For each node pair, we compute a percentage of agreement and a percentage of disagreements as shown in ~\ref{eqn:per_agreement}. The plots show the kernel density estimation (KDE) of each percentage for ``inter-party'' and ``intra-party'' edges. The figure on the right refers to the percentage of disagreements (negative edges) of the House of Representatives of 78th U.S. Congress, and the figure on the right refers to the percentage of agreement (positive edges) of the House of Representatives of 82nd U.S. Congress.
}
  \label{fig:Threshold}
\end{figure}

In order to keep only the edges that represent a clear positive or negative relationship between two politicians, two thresholds are computed for each Congress: one for the positive edges and one for the negative edges. The threshold for the positive edges is defined as the crossing point between the  kernel density estimation (KDE) of the percentage of votes in agreement of the ``intra-party'' and ``inter-party'' edges. The threshold for negative edges is computed similarly. An example is shown in Figure~\ref{fig:Threshold}. In ~\cite{plosone}, the authors apply a similar approach to the same dataset in terms of network homophily, without considering structural balance.

The yellow areas of Figure~\ref{fig:Threshold} are respectively the KDE of the distributions of negative ``inter-party'' edges (on the left) and positive ``inter-party'' edges (on the right). A positive ``inter-party'' edge represents the case when a politician votes in the same way as another politician of a different party. This is a type of frustrated edge, i.e., an edge that reduces polarization by reducing the partial structural balance of the network.

The blue areas in Figure~\ref{fig:Threshold} are respectively the KDE of distributions of negative ``intra-party'' edges and positive ``intra-party'' edges. The edges labeled as ``intra-party'' connect two politicians of the same party, so a negative ``intra-party'' edge is an example of a frustrated edge, while a negative one is not. 

The red vertical lines in Figure~\ref{fig:Threshold} represent the thresholds for positive and negative edges of the House of Representatives of the 78th and 82th U.S. Congresses.

Considering all the Congresses, for the House of Representatives there are a total of $3,436$ unique nodes, $1,984,618$ ``inter-party'' edges, and $2,055,660$ ``intra-party'' edges. For the U.S. Senate, there are a total of $707$ unique nodes, $107,843$ ``inter-party'' edges, and $108,513$ ``intra-party'' edges.

\section{Results}
\label{sec:results}
The first step in the study of the dynamics of polarization in the U.S. Congress is the analysis of network homophily. By construction, each node has a membership cluster, i.e., its party; in a signed network, network homophily is manifested by positive edges between nodes of the same party, and negative edges between nodes of different parties. The dynamics of homophily over time can be understood by comparing the frequency distributions of agreement and disagreement percentages. The overlap between the two kernel density estimation (red area in Figure \ref{fig:Threshold}) shows that i) the number of votes in agreement with the affiliation party and the number of votes in agreement with the opposing party are comparable (Figure \ref{fig:Threshold} on the left); ii) the number of votes in disagreement with the affiliation party and those in disagreement with the opposing party are also comparable (Figure \ref{fig:Threshold} on the right). This overlap can be interpreted as a tendency of politicians to reduce polarization by challenging their own party and cooperating with the opposing party, thus reducing the network homophily.

\begin{figure}[!tbp]
  \centering
  \captionsetup{width=.92\linewidth}
  \subfloat{\includegraphics[width=0.47\textwidth]{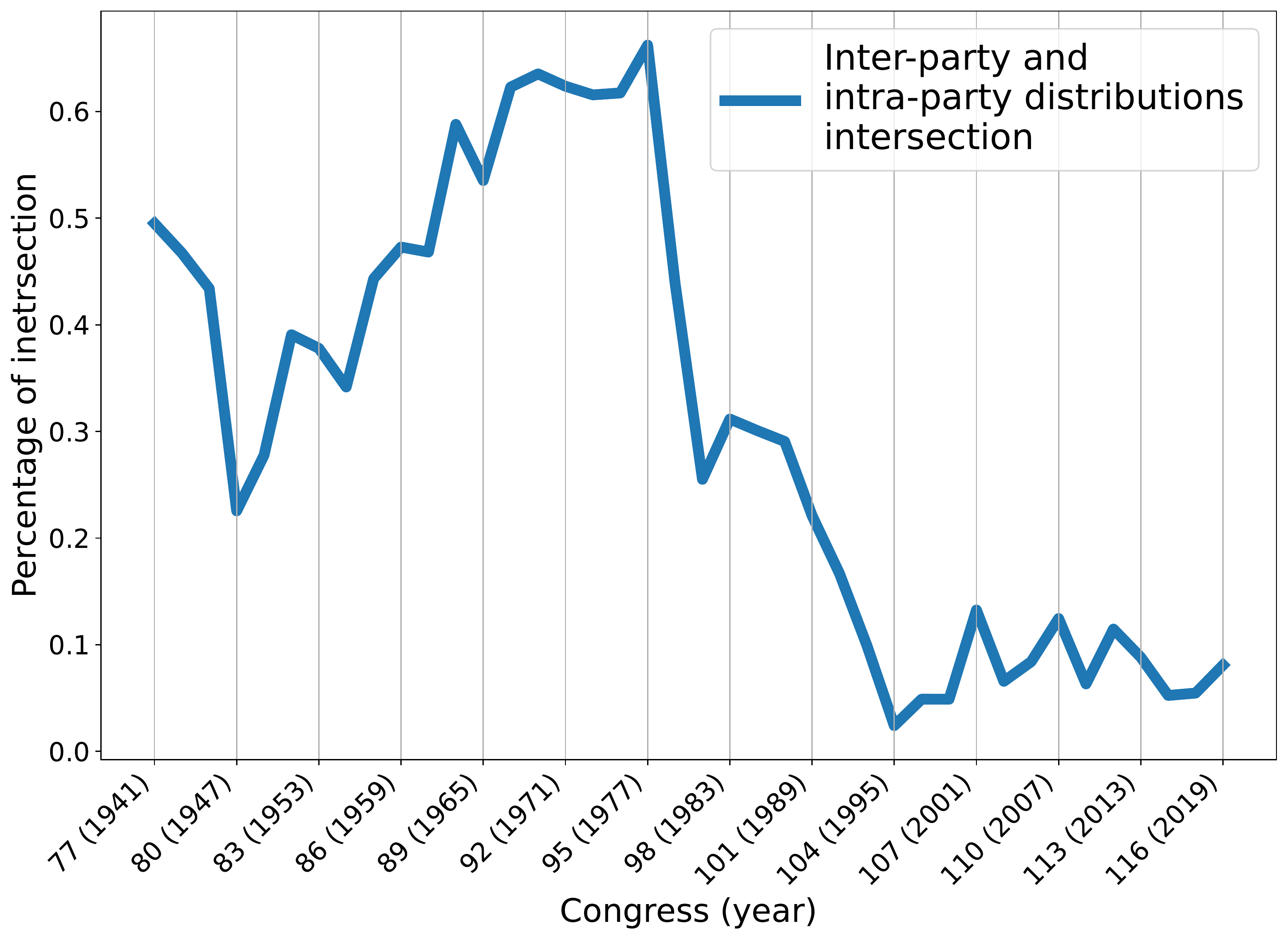}}
  \hfill
  \subfloat{\includegraphics[width=0.47\textwidth]{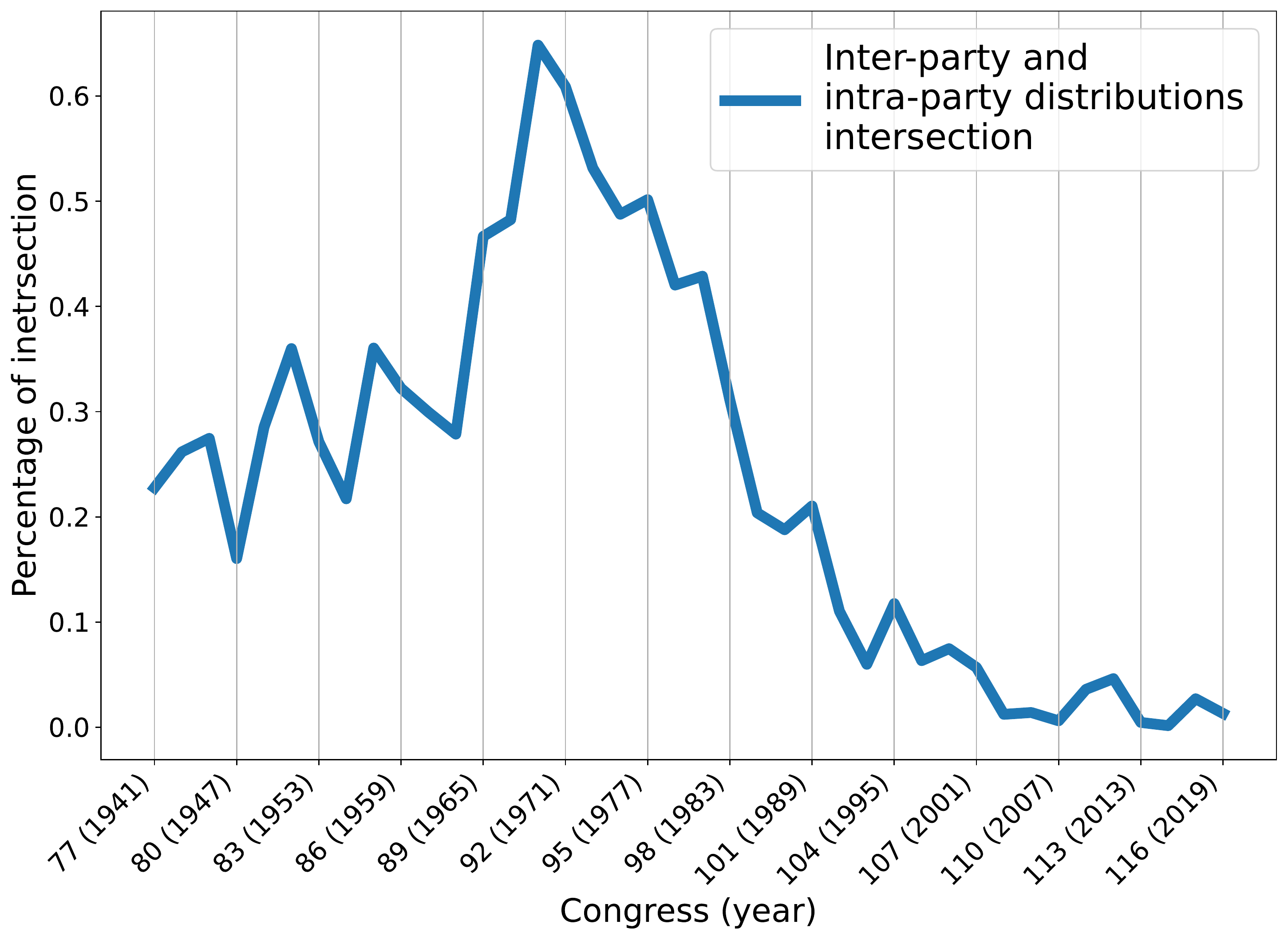}}
  \caption{Overlap between ``inter-party'' and ``intra-party'' positive edges frequency distributions over the congresses (U.S. Senate on the left and House of the Representatives on the right).}
  \label{fig:intersection}
\end{figure}

Figure~\ref{fig:intersection} shows the variation of this overlap over time, from Congress 77 to Congress 116, for the House of Representatives and the U.S. Senate. The overlap between the frequency distribution of ``inter-party'' edges and ``intra-party'' edges tends to decrease dramatically over time. This suggests an increase in polarization and a decrease in cooperation among politicians. In the next Section we verify that the increase in network homophily over time is consistent with an increase in structural balance.

Sections~\ref{ssec:frustrationalgeb} and~\ref{ssec:vizcongress} address polarization in terms of balance, both quantitatively and visually. In Section~\ref{ssec:manchin}, we show how decomposing the eigenvalues of the Laplacian matrix can bring the analysis to a level of detail not previously available in the U.S. Congressional polarization literature. Finally, Section~\ref{ssec:performance} shows the computational advantages of algebraic conflict, both with respect to the case study of the American Congress and with respect to artificially created networks. Emphasis is placed on the relationship between the percentage of frustrating edges and the computational cost of both the frustration index and the algebraic conflict.

\subsection{Frustration index and algebraic conflict}
\label{ssec:frustrationalgeb}

In order to study the dynamics of polarization through the analysis of partial structural balance, we focus on two metrics: frustration index and algebraic conflict (see section~\ref{sec:methods}).

\begin{figure}
\centering
    \captionsetup{width=.92\linewidth}
    \subfloat{\includegraphics[width=0.49\textwidth]{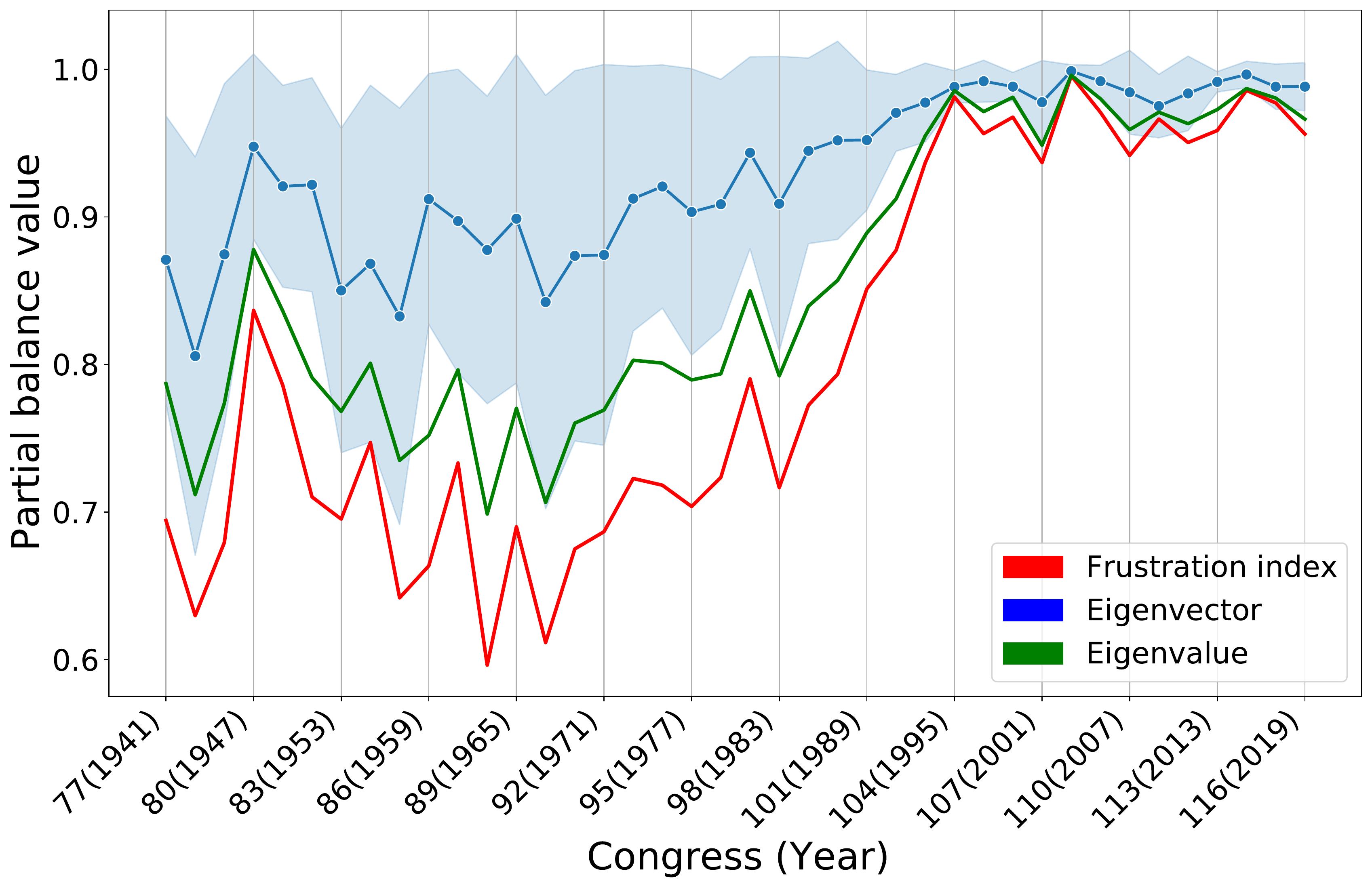}\label{fig:senate_frustration_eigen}}\hskip1ex
    \subfloat{\includegraphics[width=0.49\textwidth]{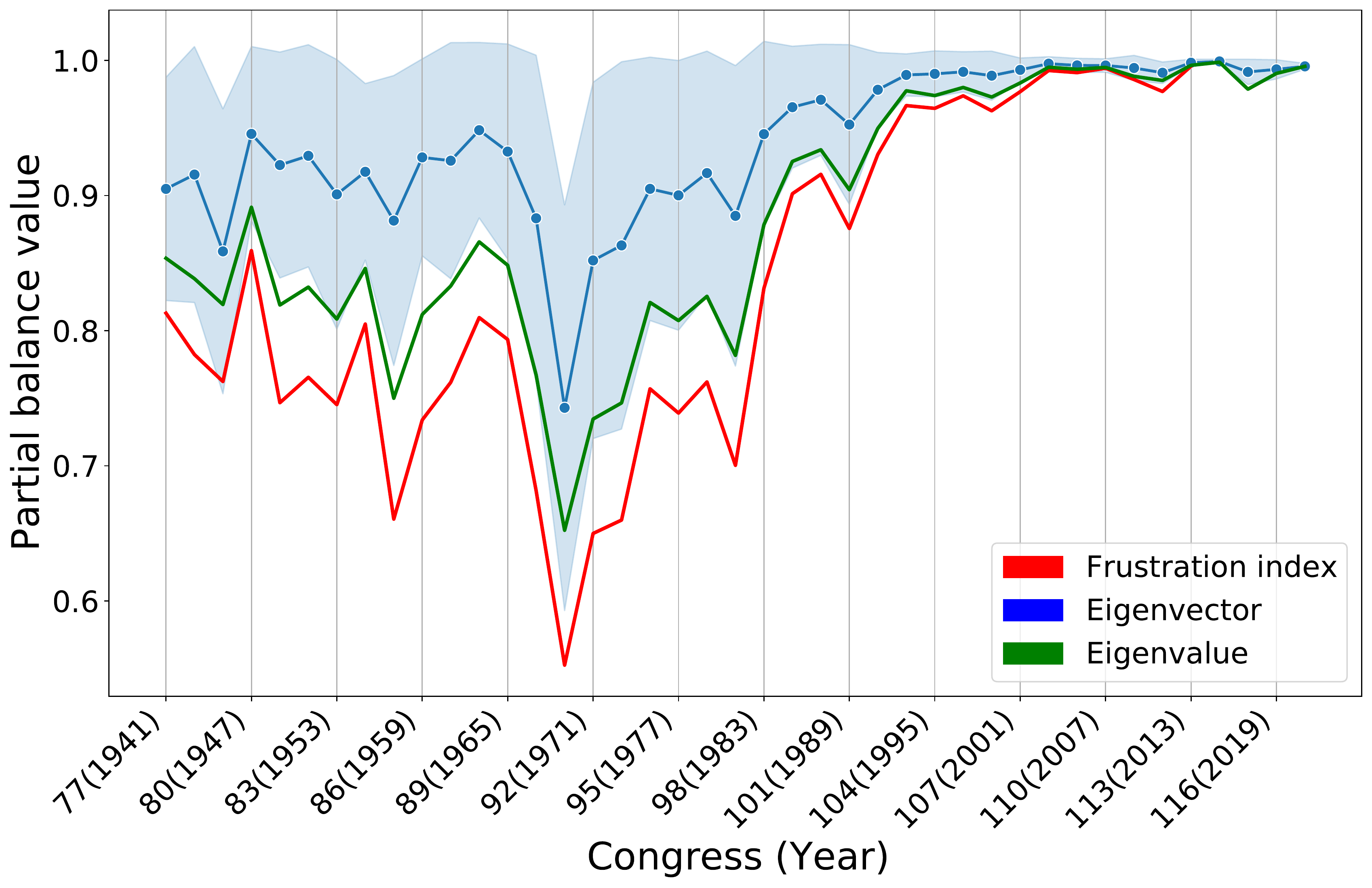}\label{fig:house_frustration_eigen}}
    \caption{Comparison across Congress of frustration index, algebraic conflict, and the values of the eigenvector corresponding to the least eigenvalue. The plot on the left refers to the U.S. Senate, the plot on the right to the House of the Representatives.}
    \label{fig:frustration_eigen}
\end{figure}

Figure \ref{fig:frustration_eigen} shows the partial structural balance of the Congresses from 1941 to 2021 computed with frustration index (red line) and algebraic conflict (green line). The blue line represents the mean of the values of the eigenvector $u_{|V|}$ corresponding to the smallest eigenvalue $\lambda_{|V|}(\Gamma)$. The blue area represents the confidence interval computed as the standard deviation of the mean of the eigenvector $u_{|V|}$.
Each element of the eigenvector $u_{|V|}$ refers to a node of the network. Nodes can be divided into two clusters using the sign of the elements of the eigenvector. The influence of each node on the structural balance of the whole network can be quantified by means of its corresponding value in $u_{|V|}$. A node not involved in frustrated edges has a corresponding eigenvector value of one. Conversely, a node with a corresponding value close to zero is involved in a process of balance reduction. Therefore, the higher the mean of the eigenvector, the more cohesive are the node clusters. A larger standard deviation indicates a greater tendency of nodes to autonomously interact with the rest of the network.

Furthermore, in Section~\ref{sec:methods} we discuss how the eigenvector of the smallest eigenvalue can be used for partitioning and positioning the nodes in a signed network visualization. In the visualizations based on the \textit{Structural-balance-viz} algorithm in Figure~\ref{fig:appendixoverview}, the reduction in the standard deviation of the eigenvector $u_{|V|}$ over time is clearly represented by the fact that the nodes become less scattered along the x-axis over time.

Pearson’s correlation is computed for all the combinations of frustration index, algebraic conflict, and the mean of the eigenvector $u_{|V|}$. For both the U.S. Senate and the House of the
Representatives, Pearson’s correlation coefficients are between $0.92$ and $0.98$, with all p-values below $0.005$.

The frustration index and the algebraic conflict increase significantly since Congress 90 (1967) for the U.S. Senate and since Congress 91 (1967) for the House of Representatives. The progressive increase in political polarization, computed in terms of both homophily and structural balance, is consistent with the literature~\cite{neal2020, Waugh2009, Ferreira2018, aref2020, Haidt}.

\subsection{Visualize the U.S. Congress}
\label{ssec:vizcongress}

The analysis of the polarization in terms of homophilia and balance can be supported by the visualization of the structural properties of the network. \textit{Structural-balance-viz}~\cite{Galimberti} algorithm, is not only computationally efficient, but also provides an at-a-glance view of network balance and which nodes are most interested in frustrated edges. 

\begin{figure}[!tbp]
	\centering
    \captionsetup{width=.92\linewidth}
    \includegraphics[width=0.7\linewidth]{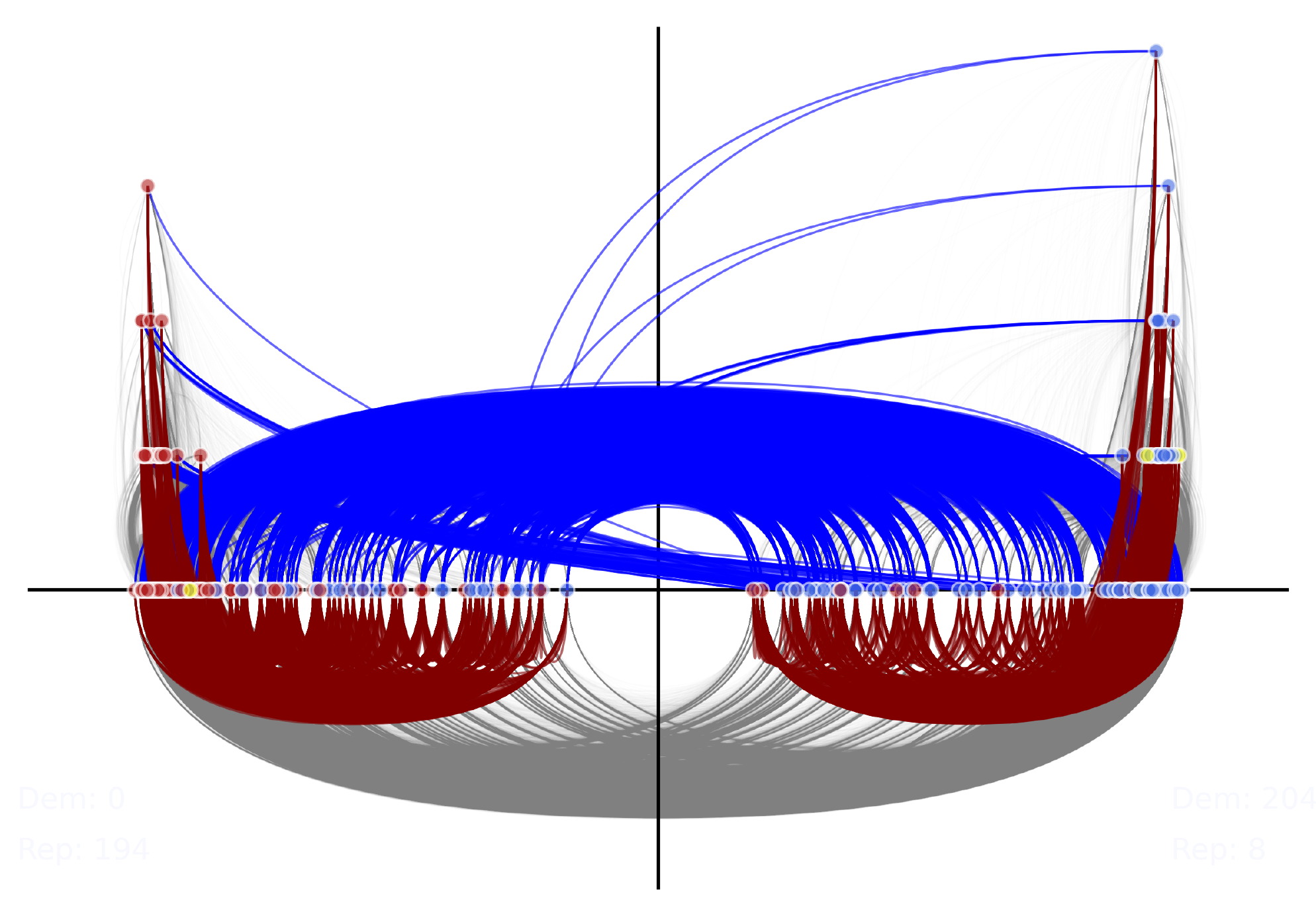}
	\caption{Network created with the \textit{Structural-balance-viz} algorithm and representing the votes of the House of Representatives of Congress 82 (1951). Non-frustrated edges ($82440$) are colored in gray, positive frustrated edges in blue ($6836$), and negative frustrated edges in red ($10605$). Nodes are color-coded according to the corresponding party: red nodes are Republicans, and blue nodes are Democrats. It is striking that a tendency to polarization is slowed down by nodes belonging to different parties whose position in the x-axis is mixed, no matter the clusters determined by the structure of the underlying signed network. The partial balance of the network partial balance computed with the algebraic conflict is $0.832$, and computed with the frustration index is $0.765$.}
	\label{fig:galimberti_house_82}
\end{figure}

Figure~\ref{fig:galimberti_house_82} shows the \textit{Structural-balance-viz} applied to the voting network of the House of Representatives of Congress 82 (1951). A large number of frustrated edges ($17441$, while e.g. Congress $117$ has only $118$ frustrated edges), both positive (blue edges) and negative (red edges), are clearly visible in the visualization. Since we are interested in highlighting the edges that reduce the partial balance of the network, we color all non-frustrated edges light gray. In addiction, the two clusters, represented by the two upper quadrants of the Cartesian plane, are heterogeneous. The nodes are highly scattered along the x-axis, although the right quadrant cluster has more blue nodes (Democratic politicians) and the left quadrant cluster has more red nodes (Republican politicians). The visualization of a highly unbalanced network is characterized by nodes scattered along the x-axis, and a significant number of frustrated edges.

\begin{figure}[!tbp]
	\centering
        \captionsetup{width=.92\linewidth}
    \includegraphics[width=0.7\linewidth]{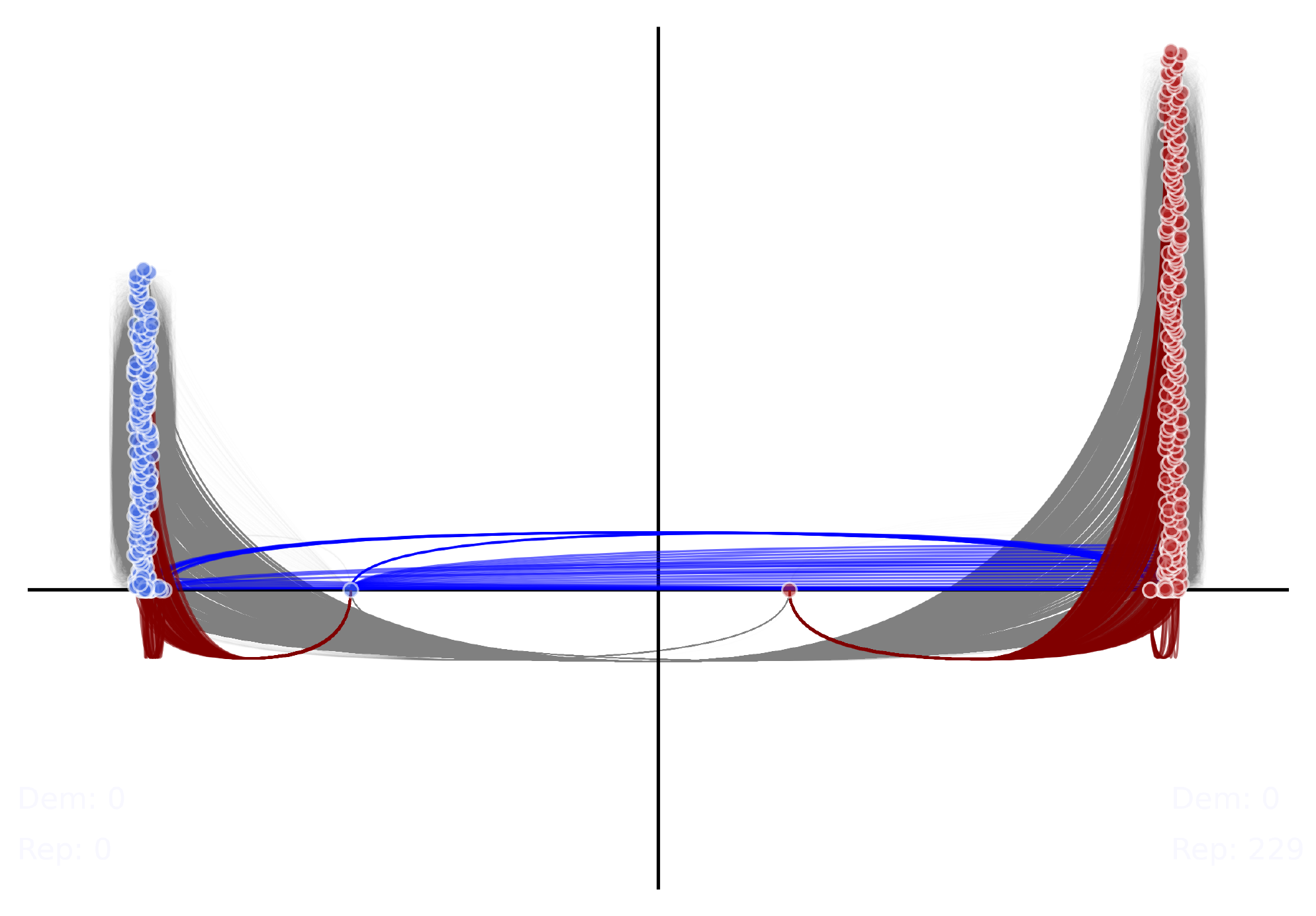}
	\caption{Network created with the \textit{Structural-balance-viz} algorithm and representing the votes of the House of Representatives of Congress 108 (2003). Non-frustrated edges are colored in gray ($96435$), positive frustrated edges in blue ($247$), and negative frustrated edges in red ($319$). Nodes are color-coded according to the corresponding party: red representatives are republicans, and blue ones are democrats. The strong signal of polarization is shown by the almost perfect match between nodes' colors and their positions at opposing sides of the x-axis.}
	\label{fig:galimberti_house_108}
\end{figure}

Figure~\ref{fig:galimberti_house_108} shows the voting network of the House of Representatives of Congress 108 (2003). In this case, it is immediately apparent that the situation is very different from that shown in Figure~\ref{fig:galimberti_house_82}: there are few positive (in blue) and negative (in red) frustrated edges, and the nodes are represented in two cohesive and homogeneous clusters. These clusters almost exactly represent the two parties, the Democrats in the right quadrant and the Republicans in the left quadrant. These are the characteristics of a highly balanced network view; the partial balance is equal to $0.995$, computed with the algebraic conflict, and to $0.992$ computed with the frustration index (see Figure~\ref{fig:frustration_eigen}). Homophily emerges more clearly in this scenario: nodes in the same party agree more often than nodes in different parties. Analogously, nodes in different parties disagree more likely than nodes in the same party.

The network would be almost totally balanced if it were not for the two nodes (a Republican and a Democrat) in the right quadrant which are spatially located closer to the origin of the x-axis, w.r.t. the nodes of their own parties.

Figure~\ref{fig:appendixoverview} shows an overview of the Structural-balance-viz algorithm applied to the voting networks of 20 congresses of the U.S. Senate between the years 1943 and 2019. Over time there is a clear progressive reduction in frustrated edges and a separation of the two clusters into increasingly polarized positions.

In the next section, we propose an analysis of the political behavior of Democratic Senator Joe Manchin over 3 Congresses (from 2015 to 2021). This fine-grained analysis is supported by the networks' visualizations created with the \textit{Structural-balance-viz} algorithm.

\subsection{Linking visualization, unbalance metric, and intra/inter-parties dynamics}
\label{ssec:manchin}

\begin{figure}[!tbp]
	\centering
        \captionsetup{width=.92\linewidth}
    \includegraphics[width=1\linewidth]{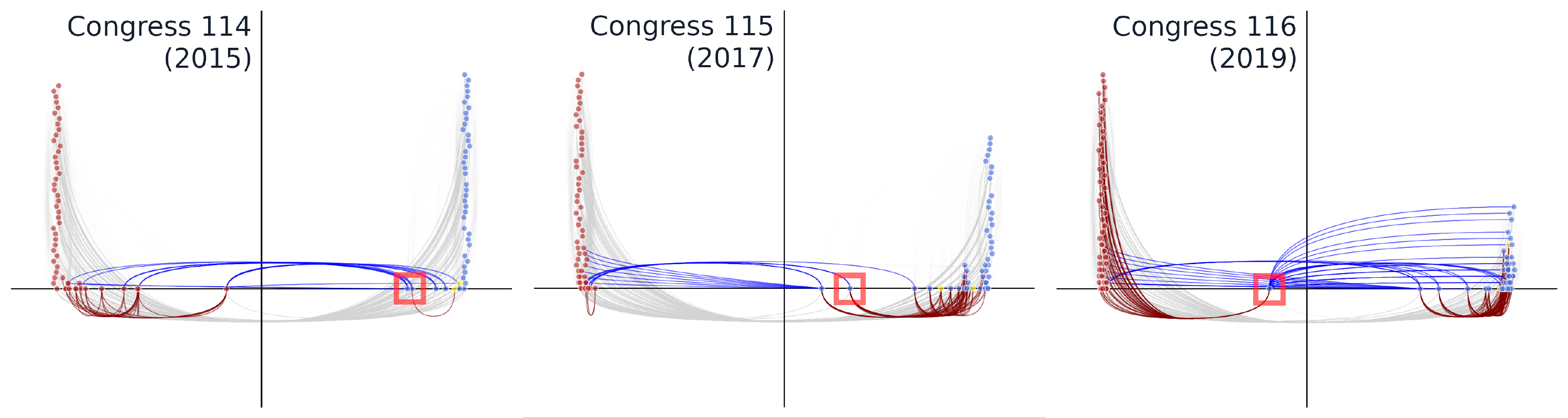}
	\caption{Networks created with the \textit{Structural-balance-viz} algorithm and representing the votes of the Senate from Congress 114 (2015) to Congress 116 (2019). The node representing Democratic Senator Joe Manchin is highlighted with a red rectangle. In only 3 congresses (6 years) he comes closer to the Republican cluster than to his party cluster.}
	\label{fig:zoomin}
\end{figure}

The possibility to study the influence of each node on the partial network balance and to follow the intra- and inter-group dynamics is one of the main advantages of studying the network balance through the spectral analysis of the Laplacian matrix. By exploiting the eigenvector $v_{|V|}$, the algorithm \textit{Structural-balance-viz} shows which are the nodes that most influence the network structure. As an example, below we show how it is possible to trace the political behavior of a Senator through the analysis of his position in three different visualizations.

The networks in Figure \ref{fig:zoomin} represent the votes of the U.S. Senate in Congresses 114, 115, and 116. The node representing Democratic Senator Joe Manchin is highlighted with a red rectangle.
His political stance has changed dramatically over the years. According to the Senate Vote Network for the 114th Congress, on average, Senator Manchin has taken positions very close to his party, with a slight tendency to cooperate with Republicans. In the next Congress, the 115th (2017-2019), the Senator moved closer to the positions of the Republican Party. In fact, the node representing Joe Manchin, while remaining in the quadrant containing the Democrat cluster, is located closer to the origin of the x-axis.

In the last Congress shown in Figure \ref{fig:zoomin}, the 116th (2019-2021), the highlighted node is in the right quadrant, meaning it is closer to the cluster of Republican nodes than to the Democratic nodes.

Senator Manchin's political transformation toward greater openness to Republicans has been a source of political debate in the newspapers since 2020\footnote{\url{https://www.theguardian.com/us-news/2021/oct/29/senate-manchin-sinema-biden-democrats-outsize-influence}}\footnote{\url{https://www.washingtonpost.com/politics/2022/01/26/manchin-sinema-democrats-legislation/}}. The analysis of the eigenvector $v_{|V|}$ of the two previous Congresses (114th and 115th) shows that this process is not unexpected, and results from a slow approach over the years to the political counterpart. Some of the interpretations justify this approach to the Republican Party with the electoral need to be re-elected. Elected in the red state of West Virginia, the Senator seems to be moving closer and closer to the demands of his potential electorate, even if they are not in line with the rest of the Democratic Party. It is also interesting that in the less polarized system observed in 1951, this behavior would have been interpreted as normal, whereas in 2020 it is clearly seen as an outlier. (Figure~\ref{fig:galimberti_house_82}).

\subsection{Unbalance metrics computational costs}
\label{ssec:performance}

The time and space complexity of computing the algebraic conflict is governed by the complexity required by the decomposition of the eigenvalues of the Laplacian matrix. In this study, for the decomposition of the eigenvalues, we use the implementation in the Python library \textit{Scipy} \footnote{\url{https://docs.scipy.org/}}. The algebraic conflict is normalized as in \ref{eq:algebconf}.

The theoretical formulation of the frustration index is NP-hard. An approximation based on simulated annealing implemented in the R library \textit{Signnet}\footnote{\url{https://cran.r-project.org/web/packages/signnet/}} is used in this study.

\begin{figure}[!tbp]
  \centering
  \captionsetup{width=.92\linewidth}
  \subfloat{\includegraphics[width=0.49\textwidth]{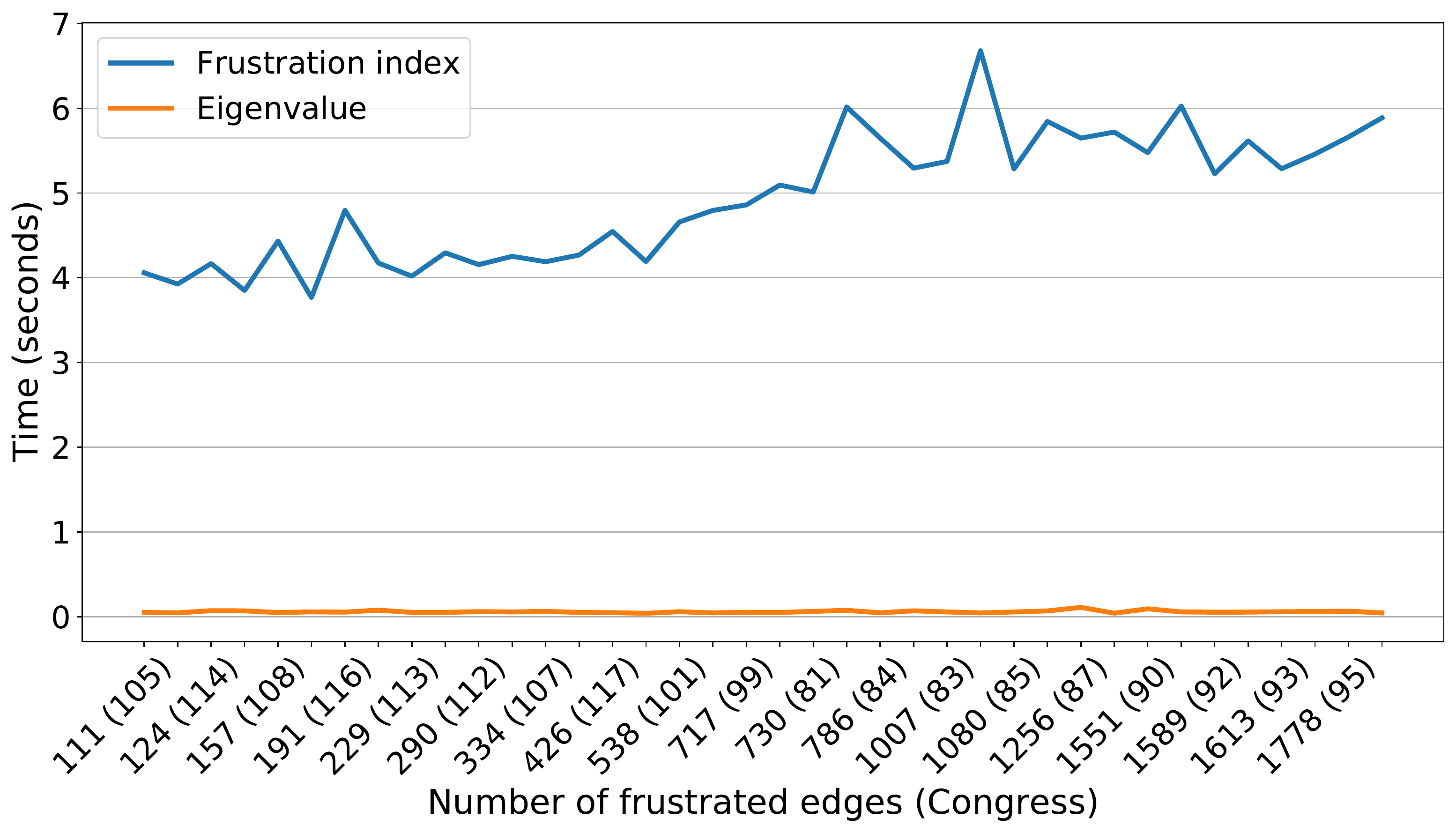}}
  \subfloat{\includegraphics[width=0.49\textwidth]{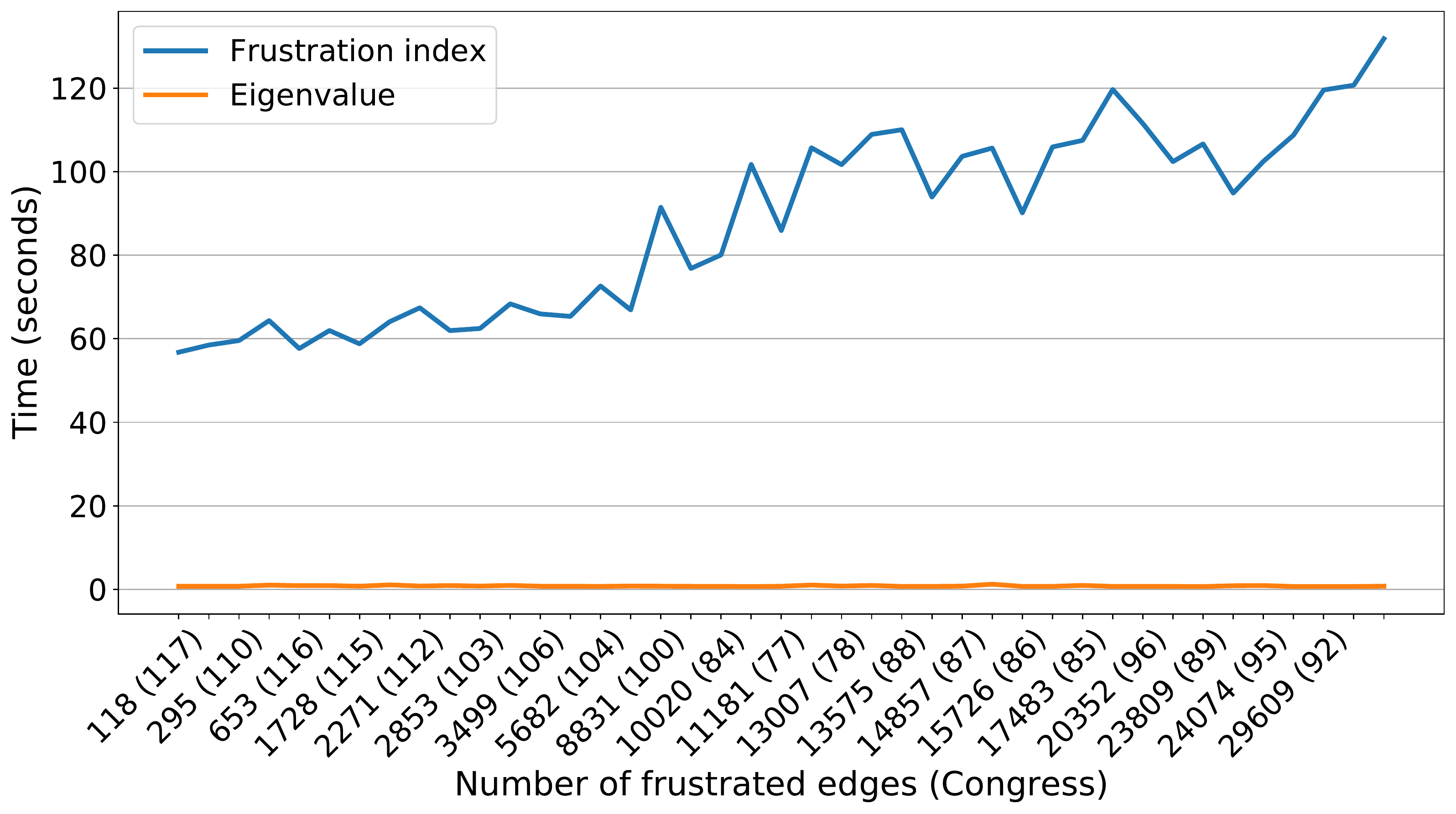}}
  \caption{The y-axis represents the computation time in seconds to compute the frustration index and the smallest eigenvalue for each congress. The networks that represent the Congresses are sorted on the x-axis according to the number of frustrated edges.
  The Figure on the left refers to the U.S. Senate, the one on the right to the House of Representatives.}
  \label{fig:compu_time}
\end{figure}

Figure \ref{fig:compu_time} shows the computation time required to compute the frustration index and the algebraic conflict for the U.S. Senate and the House of Representatives. The computations are performed on a computer with 8 GB of RAM and the Intel Core i5-6200U as CPU. The computation time for the frustration index is orders of magnitude greater than the computation time for the algebraic conflict.

Furthermore, the computation time required to compute the frustration index increases as the number of frustrated edges in a network increases (on the x-axis the networks are sorted by the number of frustrated edges). 
Since the frustration index is defined as the minimum number of edges whose removal results in balance, it is clear therefore that its computation time depends on the number of frustrated edges to remove.
The same phenomenon is not observed for the computation of the smallest eigenvalue: eigenvalue decomposition of the Laplacian matrix is independent of the edge types.

\begin{figure}[!tbp]
  \centering
  \captionsetup{width=.92\linewidth}
  \subfloat{\includegraphics[width=0.47\textwidth]{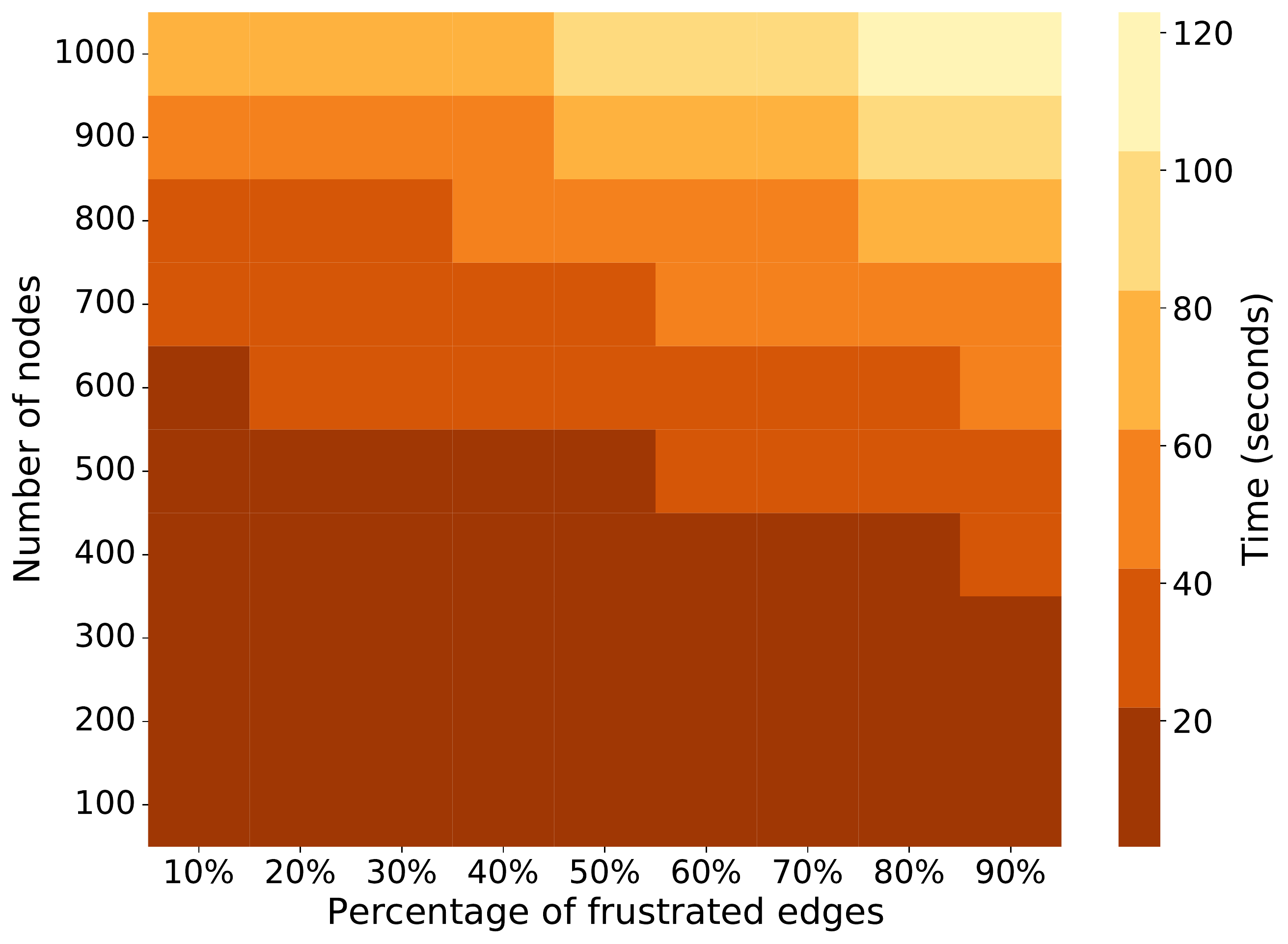}}
  \hfill
  \subfloat{\includegraphics[width=0.47\textwidth]{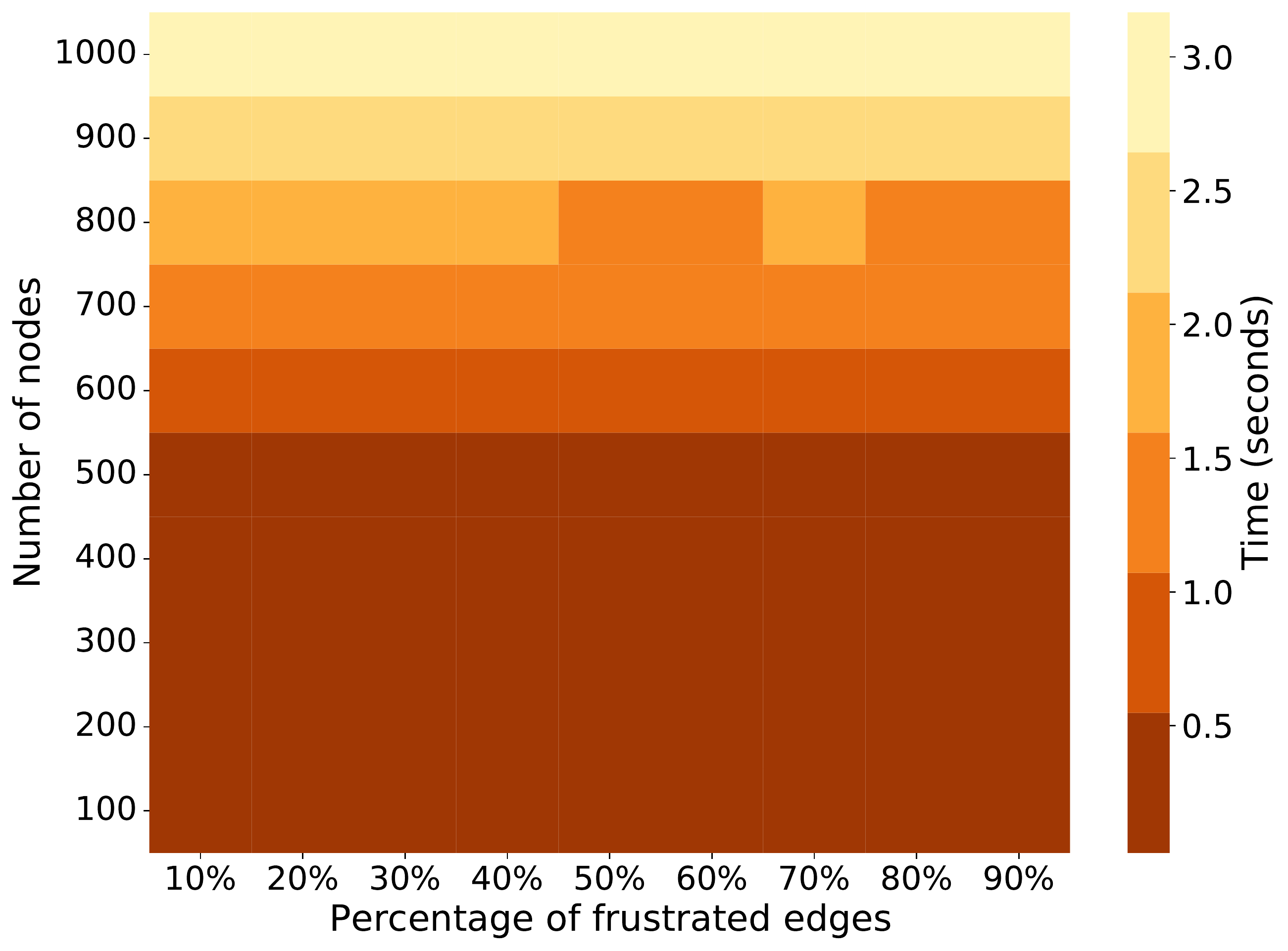}}
  \caption{Each cell of the heatmap shows the computation time in seconds required for computing the frustration index (on the left) and the eigenvalue decomposition of the Laplacian matrix (on the right). Note the difference in scale between the two heatmaps.}
  \label{fig:compu_time_heatmap}
\end{figure}

In order to generalize this concept, we generate ten random signed networks each composed of an increasing number of nodes, from $100$ to $1000$. For each network we vary the number of frustrated edges, from $10\%$ of the total number of edges, up to $90\%$.

The process of creating a random signed network is shown below:
\begin{enumerate}
\setlength\itemsep{0em}
\item Two different Erdős-Rényi networks are created (named A and B respectively).
\item Each node of the two networks is assigned an attribute based on the network it belongs to.
\item Networks A and B are merged, and edges are added between the nodes of networks A and B.
\item Each edge connecting two nodes of the same network is assigned the positive sign. Each edge between network A and network B is assigned the negative sign.
\item The network is now perfectly balanced. Based on the percentage of frustrated edges out of the total, some edges between A and B have their sign changed to positive (positive frustrated edges), and some edges inside A and inside B have their sign changed to negative (negative frustrated edges).
\end{enumerate} 

We compute both the frustration index and the smallest eigenvalue to verify that only the former has a computation time influenced by the number of frustrated edges. The results of the computation times required for each of the network configurations are shown in Figure~\ref{fig:compu_time_heatmap}, with the frustration index on the left and the eigenvalue decomposition of the Laplacian matrix on the right. For both indices, the number of nodes in the network has an effect on the computation time. However, only the computation of the frustration index is significantly affected by the percentage of frustrated edges in the network.

\section{Conclusion and future work}
\label{conclusion}

Signed networks are a powerful model for representing relationships between individuals. Community detection, as a consequence of network homophily by attribute, and partial balance computation offer the possibility to analyze conflicts and polarization between communities in a signed network of relationships. The analytical pipeline we present is based on the synergy of methods known in the literature for the study of partial structural balance. The advantages of applying this pipeline are numerous and significant: i) it offers results comparable to those obtainable with an approach based on the frustration index (Figure~\ref{fig:frustration_eigen}); ii) it allows to visualize and analyze in fine grain the influence of every single node on the structural balance of the entire network (Figures~\ref{fig:zoomin}); iii) it is very advantageous from a computational point of view (Figures~\ref{fig:compu_time}).\\
The application of this approach to the study of political polarization in the U.S. Congress has revealed implications missing from previous studies in the literature~\cite{neal2020, plosone}. This is exemplified by the case of Senator Manchin. Although it has only been in the newspapers in recent years, we show and quantify that his much-discussed political transformation began much earlier. We examine the intra-group dynamics of the senators and how this affected the partial balance of the networks over three Congresses.

In addition, since many researchers in the field of complex networks use Python as a programming language, a library for the analysis and visualization of the structural balance of the signed network is being built and released.\footnote{\url{https://github.com/alfonsosemeraro/draw_signed_networkx}}.

\appendix

\begin{figure}[!tbp]
	\centering
        \captionsetup{width=.92\linewidth}
    \includegraphics[width=1\linewidth]{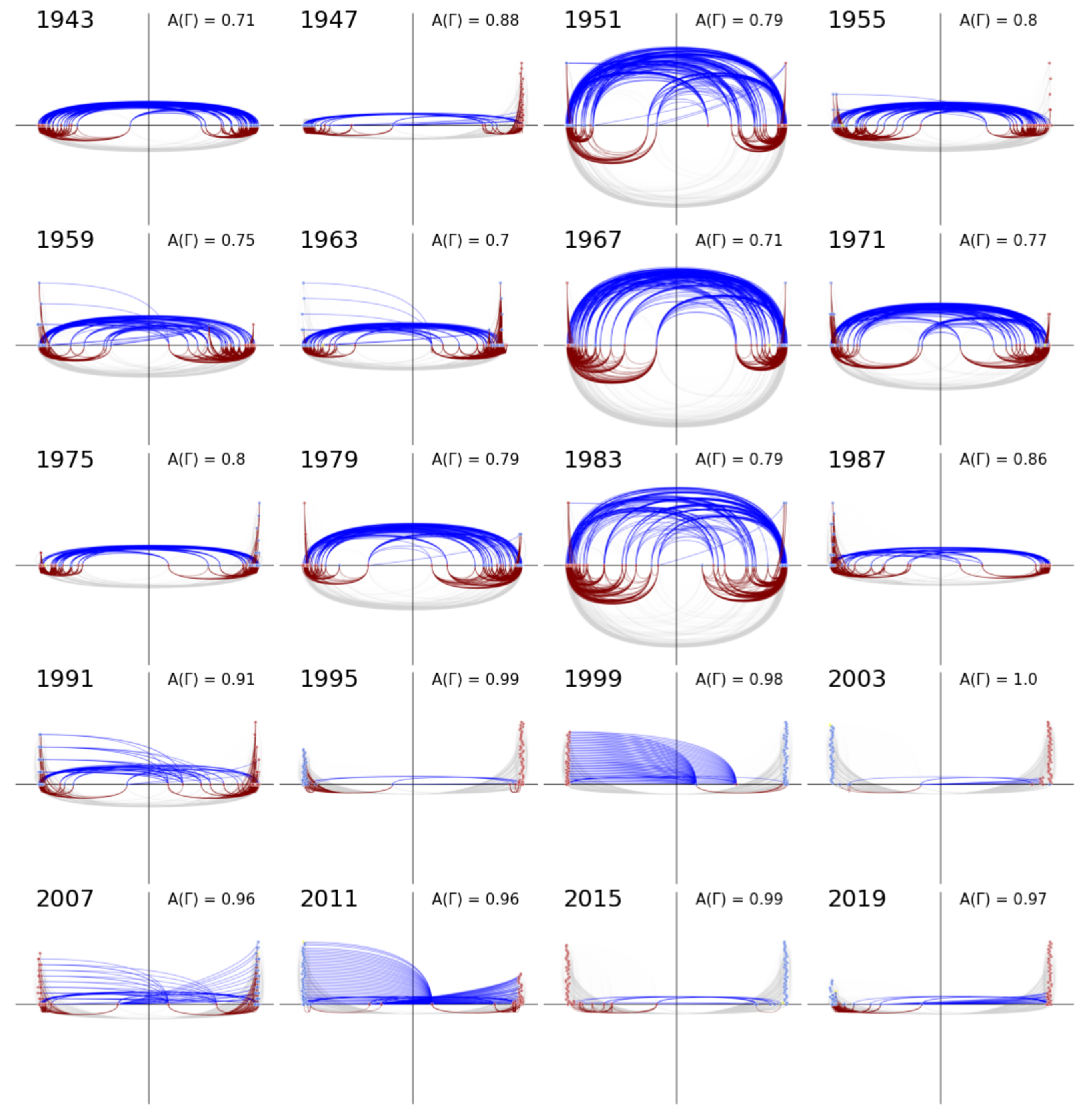}
	\caption{Structural-balance-viz algorithm applied to the voting networks of 20 congresses of the U.S. Senate between the years 1943 and 2019.}
	\label{fig:appendixoverview}
\end{figure}

\bibliographystyle{unsrt}  
\bibliography{references}

\begin{thebibliography}{10}

\bibitem{plosone}
Clio Andris, David Lee, Marcus~J. Hamilton, Mauro Martino, Christian~E.
  Gunning, and John~Armistead Selden.
\newblock The rise of partisanship and super-cooperators in the u.s. house of
  representatives.
\newblock {\em PLOS ONE}, 10:1--14, 04 2015.

\bibitem{aref2020}
Samin Aref and Zachary Neal.
\newblock Detecting coalitions by optimally partitioning signed networks of
  political collaboration.
\newblock {\em Scientific reports}, 10(1):1--10, 2020.

\bibitem{Galimberti}
Edoardo Galimberti, Chiara Madeddu, Francesco Bonchi, and Giancarlo Ruffo.
\newblock Visualizing structural balance in signed networks.
\newblock In Hocine Cherifi, Sabrina Gaito, Jos{\'e}~Fernendo Mendes, Esteban
  Moro, and Luis~Mateus Rocha, editors, {\em Complex Networks and Their
  Applications VIII}, pages 53--65, Cham, 2020. Springer International
  Publishing.

\bibitem{Aref2019}
Samin Aref and Mark~C Wilson.
\newblock Measuring partial balance in signed networks.
\newblock {\em Journal of Complex Networks}, 6(4):566--595, 09 2017.

\bibitem{Conover_Ratkiewicz_Francisco_Goncalves_Menczer_Flammini_2021}
Michael Conover, Jacob Ratkiewicz, Matthew Francisco, Bruno Goncalves, Filippo
  Menczer, and Alessandro Flammini.
\newblock Political polarization on twitter.
\newblock In {\em Proceedings of the International AAAI Conference on Web and
  Social Media}, pages 89--96, Aug. 2011.

\bibitem{sunstein2001}
Cass~R. Sunstein.
\newblock {\em Echo chambers : Bush v. Gore, impeachment, and beyond}.
\newblock Princeton University Press, 2001.

\bibitem{pariser2011}
Eli Pariser.
\newblock {\em The Filter bubble: What the Internet is hiding from you}.
\newblock Penguin, 2011.

\bibitem{lazer2018}
David M.~J. Lazer, Matthew~A. Baum, Yochai Benkler, Adam~J. Berinsky, Kelly~M.
  Greenhill, Filippo Menczer, Miriam~J. Metzger, Brendan Nyhan, Gordon
  Pennycook, David Rothschild, Michael Schudson, Steven~A. Sloman, Cass~R.
  Sunstein, Emily~A. Thorson, Duncan~J. Watts, and Jonathan~L. Zittrain.
\newblock The science of fake news.
\newblock {\em Science}, 359(6380):1094--1096, 2018.

\bibitem{Keuchenius2021}
Anna Keuchenius, Petter Törnberg, and Justus Uitermark.
\newblock Why it is important to consider negative ties when studying polarized
  debates: A signed network analysis of a dutch cultural controversy on
  twitter.
\newblock {\em PLOS ONE}, 16(8):1--23, 08 2021.

\bibitem{Kunegis2}
Jérôme Kunegis, Stephan Schmidt, Andreas Lommatzsch, Jürgen Lerner, Ernesto
  W.~De Luca, and Sahin Albayrak.
\newblock Spectral analysis of signed graphs for clustering, prediction and
  visualization.
\newblock In {\em Proc. of the 2010 SIAM International Conference on Data
  Mining (SDM)}, pages 559--570, 2010.

\bibitem{Knoke2008}
David Knoke and Song Yang.
\newblock {\em Social Network Analysis}.
\newblock SAGE Publications, Inc., Thousand Oaks, California, 2008.

\bibitem{Aggarwal}
Charu~C. Aggarwal.
\newblock An introduction to social network data analytics.
\newblock In {\em Social Network Data Analytics}, 2011.

\bibitem{Tang2017}
Jiliang Tang, Yi~Chang, Charu Aggarwal, and Huan Liu.
\newblock A survey of signed network mining in social media.
\newblock {\em ACM Computing Survey}, 49(3), aug 2016.

\bibitem{Heider}
Fritz Heider.
\newblock Attitudes and cognitive organization.
\newblock {\em The Journal of Psychology}, 21(1):107--112, 1946.

\bibitem{Cart}
Dorwin Cartwright and Frank Harary.
\newblock Structural balance: A generalization of heider's theory.
\newblock {\em Psychological Review}, 63(5):277--293, 1956.

\bibitem{Leskovec}
Jure Leskovec, Daniel Huttenlocher, and Jon Kleinberg.
\newblock Signed networks in social media.
\newblock In {\em Proceedings of the SIGCHI Conference on Human Factors in
  Computing Systems}, CHI '10, page 1361–1370, New York, NY, USA, 2010.
  Association for Computing Machinery.

\bibitem{Kunegis}
Jérôme Kunegis.
\newblock Applications of structural balance in signed social networks, 2014.

\bibitem{Doreian}
Patrick Doreian and Andrej Mrvar.
\newblock Structural balance and signed international relations: Joss.
\newblock {\em Journal of Social Structure}, 16:1--49, 2015.

\bibitem{Saiz}
Hugo Saiz, Jesús Gómez-Gardeñes, Paloma Nuche, Andrea Girón, Yolanda Pueyo,
  and Concepción~L. Alados.
\newblock Evidence of structural balance in spatial ecological networks.
\newblock {\em Ecography}, 40(6):733--741, 2017.

\bibitem{Leskovec2010a}
Jure Leskovec, Daniel Huttenlocher, and Jon Kleinberg.
\newblock Signed networks in social media.
\newblock In {\em Proceedings of the SIGCHI Conference on Human Factors in
  Computing Systems}, CHI '10, page 1361–1370, New York, NY, USA, 2010.
  Association for Computing Machinery.

\bibitem{Leskovec2010b}
Jure Leskovec, Daniel Huttenlocher, and Jon Kleinberg.
\newblock Predicting positive and negative links in online social networks.
\newblock In {\em WWW '10: Proceedings of the 19th international conference on
  World wide web}, WWW '10, page 641–650, New York, NY, USA, 2010.
  Association for Computing Machinery.

\bibitem{Victor2011}
Patricia Victor, Chris Cornelis, Martine De~Cock, and Ankur Teredesai.
\newblock Trust- and distrust-based recommendations for controversial reviews.
\newblock {\em IEEE Intelligent Systems}, 26:48--55, 01 2011.

\bibitem{Ma2009}
Hao Ma, Michael~R. Lyu, and Irwin King.
\newblock Learning to recommend with trust and distrust relationships.
\newblock In {\em Proceedings of the Third ACM Conference on Recommender
  Systems}, RecSys '09, page 189–196, New York, NY, USA, 2009. Association
  for Computing Machinery.

\bibitem{Chen2020}
Jianrui Chen, Lidan Wei, Liji U, and Fei Hao.
\newblock A temporal recommendation mechanism based on signed network of user
  interest changes.
\newblock {\em IEEE Systems Journal}, 14(1):244--252, 2020.

\bibitem{Hua_Liu2020}
Hua Liu, Cunquan Qu, Yawei Niu, and Guanghui Wang.
\newblock The evolution of structural balance in time-varying signed networks.
\newblock {\em Future Generation Computer Systems}, 102:403--408, 2020.

\bibitem{PorcoA2015}
Aldo Porco, Andreas Kaltenbrunner, and Vicenç Gómez.
\newblock Low-rank approximations for predicting voting behaviour.
\newblock In {\em Proc. of NIPS 2015 Workshop: Networks in the Social and
  Information Sciences}, 01 2015.

\bibitem{Intal2021}
Carla Intal and Taha Yasseri.
\newblock Dissent and rebellion in the house of commons: a social network
  analysis of brexit-related divisions in the 57th parliament.
\newblock {\em Applied Network Science}, 6:1--12, 2021.

\bibitem{Bonacich2007}
Phillip Bonacich.
\newblock Some unique properties of eigenvector centrality.
\newblock {\em Social Networks}, 29(4):555--564, 2007.

\bibitem{neal2020}
Zachary~P. Neal.
\newblock A sign of the times? weak and strong polarization in the u.s.
  congress, 1973–2016.
\newblock {\em Social Networks}, 60:103--112, 2020.
\newblock Social Network Research on Negative Ties and Signed Graphs.

\bibitem{Arinik2017}
Nejat Arinik, Rosa Figueiredo, and Vincent Labatut.
\newblock Signed graph analysis for the interpretation of voting behavior.
\newblock In {\em Proceedings of the Workshop Papers of i-Know 2017}, 10 2017.

\bibitem{Mendonca2015}
Israel Mendonça, Rosa Figueiredo, Vincent Labatut, and Philippe Michelon.
\newblock Relevance of negative links in graph partitioning: A case study using
  votes from the european parliament.
\newblock In {\em 2015 Second European Network Intelligence Conference}, pages
  122--129, 2015.

\bibitem{Waugh2009}
Andrew~Scott Waugh, Liuyi Pei, James~H. Fowler, Peter~J. Mucha, and Mason~A.
  Porter.
\newblock Party polarization in congress: A network science approach, 2009.

\bibitem{Dana2013}
Dana~R. Fisher, Joseph Waggle, and Philip Leifeld.
\newblock Where does political polarization come from? locating polarization
  within the u.s. climate change debate.
\newblock {\em American Behavioral Scientist}, 57(1):70--92, 2013.

\bibitem{Ferreira2018}
Carlos Ferreira, Breno Matos, and Jussara Almeira.
\newblock Analyzing dynamic ideological communities in congressional voting
  networks.
\newblock In {\em Social Informatics. SocInfo 2018. LNCS Vol. 11185}, pages
  257--273. Springer, 09 2018.

\bibitem{Facchetti}
Giuseppe Facchetti, Giovanni Iacono, and Claudio Altafini.
\newblock Computing global structural balance in large-scale signed social
  networks.
\newblock {\em Proceedings of the National Academy of Sciences},
  108(52):20953--20958, 2011.

\bibitem{Estrada}
Ernesto Estrada and Michele Benzi.
\newblock Walk-based measure of balance in signed networks: Detecting lack of
  balance in social networks.
\newblock {\em Physical Review E}, 90:042802, Oct 2014.

\bibitem{Singh_2017}
Ranveer Singh and Bibhas Adhikari.
\newblock Measuring the balance of signed networks and its application to sign
  prediction.
\newblock {\em Journal of Statistical Mechanics: Theory and Experiment},
  2017(6):063302, jun 2017.

\bibitem{Kirkley}
Alec Kirkley, George~T. Cantwell, and M.~E.~J. Newman.
\newblock Balance in signed networks.
\newblock {\em Physical Review E}, 99:012320, Jan 2019.

\bibitem{Zaslavsky2013MatricesIT}
Thomas Zaslavsky.
\newblock Matrices in the theory of signed simple graphs.
\newblock {\em arXiv: Combinatorics}, 2013.

\bibitem{DAS2004715}
K.Ch. Das.
\newblock The laplacian spectrum of a graph.
\newblock {\em Computers \& Mathematics with Applications}, 48(5):715--724,
  2004.

\bibitem{Marsden2013EIGENVALUESOT}
Ann Marsden.
\newblock Eigenvalues of the laplacian and their relationship to the
  connectedness, 2013.

\bibitem{barahona1983}
Francisco Barahona.
\newblock The max-cut problem on graphs not contractible to k5.
\newblock {\em Operations Research Letters}, 2(3):107--111, 1983.

\bibitem{Belkin}
Mikhail Belkin and Partha Niyogi.
\newblock Laplacian eigenmaps and spectral techniques for embedding and
  clustering.
\newblock In T.~Dietterich, S.~Becker, and Z.~Ghahramani, editors, {\em
  Advances in Neural Information Processing Systems}, volume~14. MIT Press,
  2001.

\bibitem{ColemanClustering}
Tom Coleman, James Saunderson, and Anthony Wirth.
\newblock Spectral clustering with inconsistent advice.
\newblock In {\em Proceedings of the 25th International Conference on Machine
  Learning}, ICML '08, page 152–159, New York, NY, USA, 2008. Association for
  Computing Machinery.

\bibitem{BELARDO2014133}
Francesco Belardo.
\newblock Balancedness and the least eigenvalue of laplacian of signed graphs.
\newblock {\em Linear Algebra and its Applications}, 446:133--147, 2014.

\bibitem{Hou2005BoundsFT}
Yaoping Hou.
\newblock Bounds for the least laplacian eigenvalue of a signed graph.
\newblock {\em Acta Mathematica Sinica}, 21:955--960, 2005.

\bibitem{Haidt}
Jonathan Haidt.
\newblock {\em The Righteous Mind: Why Good People Are Divided by Politics and
  Religion}.
\newblock Vintage Books, New York, New York, United States, 2008.

\end{thebibliography}

\end{document}